\documentclass[preprint,12pt]{elsarticle}
\usepackage{graphicx}
\usepackage{amssymb}

\journal{Physics Reports}

\begin{document}

\begin{frontmatter}

\title{The Astrophysics of the Intracluster Plasma}

\author[1,2]{Alfonso Cavaliere}\sep\author[1,3]{Andrea Lapi}

\address[1]{Univ. `Tor Vergata', Via Ricerca Scientifica 1, 00133 Roma, Italy.}
\address[2]{INAF, Osservatorio Astronomico di Roma, via Frascati 33, 00040
Monteporzio, Italy.}
\address[3]{SISSA, Via Bonomea 265, 34136 Trieste, Italy.}

\begin{abstract}
Since 1971 observations in X rays of thousands galaxy clusters have
uncovered huge amounts of hot baryons filling up the deep gravitational
potential wells provided by dark matter (DM) halos with sizes of millions
light-years and masses of some $10^{15}\, M_{\odot}$. At temperatures $T
\sim 10^8$ K and with average densities of $n \sim 1$ particle per liter,
such baryons add up to some $10^{14}\, M_{\odot}$. With the neutralizing
electrons, they constitute the best proton-electron \textit{plasma} in the
Universe (whence the apt name Intra Cluster Plasma, ICP), one where the
thermal energy per particle overwhelms the average electron-proton Coulomb
interactions by extralarge factors of order $10^{12}$. The ICP shines in X
rays by thermal bremsstrahlung radiation, with powers up to several
$10^{45}$ erg s$^{-1}$ equivalent to some $10^{11}$ solar luminosities.

The first observations were soon confirmed in X rays by the detection of
high excitation emission lines, and in the radio band by studies of
streamlined radiogalaxies moving through the ICP. Later on they were nailed
down by the first measurements in microwaves of the Sunyaev-Zel'dovich
effect, i.e., the inverse Compton upscattering of cold cosmic background
photons at $T_{\rm cmb}\approx 2.73$ K off the hot ICP electrons at $k_B T
\sim 5$ keV.

A key physical feature of the ICP is constituted by its good local thermal
equilibrium, and by its overall hydrostatic condition in the DM wells,
modulated by entropy. The latter is set up in the cluster center by the
initial halo collapse, and is progressively added at the outgrowing cluster
boundary by standing shocks in the supersonic flow of intergalactic gas
into the DM potential wells. Such physical conditions are amenable to
detailed modeling. We review here these entropy-based models and discuss
their outcomes and predictions concerning the ICP observables in X rays and
in microwaves, as well as the underlying DM parameters. These quantitative
outcomes highlight the tight relationship between the detailed ICP profiles
and the cosmological evolution of the containing DM potential wells. The
results also provide the simplest baseline for disentangling a number of
additional and intriguing physical processes superposed to the general
equilibrium.

The present \emph{Report} is focused on the ICP physics as specifically
driven by the two-stage evolution of the containing DM halos. We
extensively discuss the basic entropy pattern established by the cluster
formation and development, and cover: the central entropy erosion produced
by radiative cooling that competes with the intermittent energy inputs
mainly due to active galactic nuclei and mergers; outer turbulent support
linked with weakening shocks and decreasing inflow through the virial
boundary, causing reduced entropy production during the late stage of DM
halo evolution; the development from high to low entropy levels throughout
a typical cluster; perturbations of the equilibrium up to outright
disruption due to deep impacts of infalling galaxy groups or collisions
with comparable companion clusters; relativistic energy distributions of
electrons accelerated during such events, producing extended radio emission
by synchrotron radiation, and contributing to non-thermal pressure support
for the ICP.

We conclude with discussing selected contributions from cluster
astrophysics to cosmology at large, and by addressing how the ICP features
and processes will constitute enticing targets for observations with the
ongoing \textsl{Planck} mission, for upcoming instrumentation like
\textsl{ALMA} and other ground-based radio observatories, and for the
next-generation of X-ray satellites from \textsl{ASTRO-H} to
\textsl{eROSITA}.
\end{abstract}

\begin{keyword}
galaxies: clusters: general \sep X rays: galaxies: clusters \sep X rays:
intracluster medium \sep cosmic background radiation
\end{keyword}

\end{frontmatter}

\section{Hot Baryons in Galaxy Clusters}

F. Zwicky championed back in 1933 the notion that the high line-of-sight
galaxy velocities of order $10^3$ km s$^{-1}$ optically observed in clusters
were to be interpreted not as bulk motions toward or outward of a chance
`constellation' of $\sim 10^3$ galaxies, but rather as $1$-D random
velocities with dispersions $\sigma_r$ in a statistically \textit{steady}
ensemble confined by gravity within a few Mpcs. The virial theorem
appropriate for an approximately spherical and homogeneous mass distribution
reads
\begin{equation}
\sigma^2_r = G\, M/5\, R~.
\end{equation}
This shows such high random velocities to require for `rich' clusters (as
defined by Abell 1958) a then outlandish binding mass around $M\sim 10^{15}\,
M_{\odot}$, far larger than the sum of all galaxies in a cluster. Now we know
this to constitute a major piece of evidence for non-baryonic dark matter
(DM), but controversy lingered down to the early 1970's, and was still echoed
in otherwise knowledgeable textbooks of the time on basic Astronomy.

As to the baryonic content, any gas at thermal equilibrium floating in the
system would feature comparably high thermal velocities, corresponding to
temperatures
\begin{equation}
k_B\, T = m_p\, \sigma^2_r/2 \approx 5~\mathrm{keV}~,
\end{equation}
$k_B$ being the Boltzmann constant and $m_p$ the proton mass. Thus it will
emit by thermal bremsstrahlung photons in a continuum out to several keVs,
smack in the middle of the X-ray band. Such an emission was indeed detected
in 1971 for a handful of nearby clusters standing out in the quick-look data
from the first X-ray satellite \textsl{Uhuru}, that by the technology of the
time was particularly sensitive in the $2-6$ keV range.

Cavaliere, Gursky \& Tucker (1971) were the first to suggest these ought to
be just the tips of a new class of extragalactic X-ray sources; similar
conditions should prevail in \textit{all} galaxy clusters and even in smaller
galaxy associations like \emph{groups} with appropriately cooler X-ray
temperatures $k_B T \lesssim 1$ keV, as indeed found later on in the 1990's
in the softer bands of \textsl{Einstein} and \textsl{ROSAT}. The above
authors stressed that the emission process would most likely involve an
extended, thermal but not necessarily isothermal intra\,cluster medium rather
than a multiplicity of non-thermal sources like active galactic nuclei as
contended for a while.

These notions were nailed down by the detection and by size measurements of a
growing number of such cluster sources from Gursky et al. (1972) to Forman et
al. (1978). They squared up with the observations in the radioband of
head-tail radiogalaxies streamlined by the ram pressure from their motion
through the intracluster medium (see Miley et al. 1972). On the other hand,
Sunyaev \& Zel'dovich (1972) were prompt to point out that the hot electrons
in the ICP are bound to cause inverse-Compton upscattering of cold cosmic
microwave background (CMB) photons crossing the clusters, so providing an
independent probe of the thermal pressure in the intracluster medium.
Finally, all controversy was ended by the first detections of high
excitation, coronal-like emission lines; these pinpointed conditions of
thermal equilibrium from a few to several keVs (see Mitchell et al. 1976,
Serlemitsos et al. 1977), and also indicated definite, subsolar metallicities
$Z\sim 0.3-0.5\, Z_{\odot}$.

On the other hand, the hot medium is bound to emit in X rays copious thermal,
optically-thin bremsstrahlung radiation, to attain from the cluster volume
$V$ bright luminosities $L_X\approx 2 \times 10^{-27}\, n^2\,\sqrt{T}\, V
\sim 10^{44-46}$ erg s$^{-1}$. These emissions enable us to probe in fine
detail the surface brightness profiles and so the number densities, in some
instances out to the virial radius $R$; these profiles turn out to peak at
$n\sim 10^{-2}$ cm$^{-3}$ in the centers, and to decline outwards by factors
of order $10^2$. The radial integration of the full profiles yields large
amounts of baryons up to masses $m \sim 10^{14}\, M_{\odot}$, implying a
baryon to DM ratio $m/M \sim 1/6$ close to the universal value (see White et
al. 1993). It is interesting to note that shortly before 1971 false X-ray
detections and upper limits had still yielded baryonic contents ranging by an
order of magnitude up and down relative to the true amount (see also Sarazin
1988 for the historical context and developments).

In fact, we now know that the DM `halo' accounts for some $6/7$ of the total
masses $M \sim 10^{13-15}\, M_{\odot}$ from poor groups to rich clusters,
with its 'collisionless' constituent particles entertaining little or no
interactions other than gravity. So it is the DM that sets the overall
gravitational wells virialized within radii $R\sim $ a few Mpcs, where all
bodies in dynamical equilibrium $-$ from whole galaxies to single particles
$-$ possess or acquire the $1$-D velocity dispersions $\sigma_r\sim 10^3$ km
s$^{-1}$ entering Eq.~(1).

The bulk of the baryons, to a fraction that happens to approach $6/7$ of
their total, is found in the \emph{diffuse} form of a hot intracluster
medium. Given its temperature well above most ionization potentials, this is
fully ionized, and is mostly comprised of protons and neutralizing electrons
close to local thermal equilibrium, with proton mean free paths $\lambda_{\rm
pp}$ ranging from a few kpcs at the center to some $10^2$ kpc in the
outskirts.

To put such numbers in a physical perspective, note that the constituent
particles floating in the DM gravitational wells must have very large
specific kinetic relative to the electrostatic energy at their mean
separation $d \sim n^{-1/3}$; in fact, the ratio $k_B T/e^2\,n^{1/3} \sim
10^{12}$ implies the latter to be by far dominated by the former energy. This
is an astounding ratio when compared with its counterparts: some $10^{3}$ in
stellar interiors, or $3\times 10^5$ in the pre-recombination Universe. It
applies despite gravity being so exceedingly feeble at a microscopic level as
to attain a mere $G\, m_p^2 /e^2 \sim 8\times 10^{-37}$ of the strength that
marks the electromagnetic interactions. This occurs because the condition
\begin{equation}
10^{12} \sim k_B\, T/e^2 \, n^{1/3} = G\, m_p^2/e^2 \, \times d/10\, R\,
\times N
\end{equation}
is dominated by the huge number $N\equiv M/m_p \sim 10^{73}$ expressing in
proton units the total DM mass with its overwhelming gravity. As a result,
the intracluster medium by far constitutes the best proton-electron
\textit{plasma} in the Universe ever, and may be aptly named IntraCluster
Plasma (henceforth ICP, including the intragroup plasma). Equivalently, the
particle number in a Debye cube is as large as $n\, \lambda_D^3\sim (k_B
T/4\pi e^2 n^{1/3})^{3/2}\sim 10^{16}$, which ensures the ICP can be very
closely treated as a single fluid.

Thus the ICP constitutes a nearly perfect gas of protons and electrons with
$3$ degrees of freedom and effective mass $\mu m _p$ in terms of the mean
molecular weight $\mu \approx 0.6$. At intermediate scales of some $10$ kpc
the protons share their momentum and energy over mean free paths
$\lambda_{\rm pp} \sim 10\, (k_B\, T/ 5 \,\mathrm{keV})^2$
$(n/10^{-3}\,\mathrm{cm}^{-3})^{-1}$ kpc, and the electrons follow suite over
some $40\, \lambda_{\rm pp}$ toward local thermal equilibrium.

In simple terms, the total thermal energy in these ICP `clouds' is up to
$10^{65}$ ergs, comparable with the fireball from a titanic $H$-bomb of some
$10^{43}$ Megaton. Given that (differently from most laboratory plasmas) the
magnetic pressure is often subdominant in the ICP, the containment of such a
cloud plainly requires a monster gravitationally binding mass of some
$10^{15}\, M_{\odot}$. By such simple considerations, the mere detection in
1971-72 of many clusters in X rays swept away at a single stroke any
lingering doubt over the `stability' of galaxy clusters and the reality of
dark halos. It may be reassuring to note that these enormous super-hot
hydrogen clouds cannot explode since their tenuous densities make nuclear
reactions rates slow even compared to the Hubble scales, by virtue of the long
mean free paths for $p-p$ reactions and the low deuterium cosmological
abundance $D/H\sim 10^{-5}$. We will see that such clouds actually result not
from explosions but rather from prolonged gravitational \textit{implosions}
of intergalactic gas along with DM.

Excellent reviews have been recently devoted to broadband descriptions of the
formation and evolution of galaxy clusters (Kravtsov \& Borgani 2012) and of
cosmology with cluster observations (Allen et al. 2011). In this
\emph{Report} we will focus on thermodynamic \emph{entropy} as the pivotal
hinge to link the ICP physics with the collapse and growth of the containing
DM halos; this not only provides a handy \emph{tool} for analysing the ICP
observables, but also yields a coherent view of cluster evolution. In
equilibrium within the gravitational potential well, the \emph{collisionless}
DM and the \emph{collisional} ICP will be distributed in different and
telling ways, worth to be measured and modeled in detail.

In the following, we will adopt the standard, flat $\Lambda$CDM cosmology
with currently accelerating expansion (cf. Hinshaw et al. 2013;
\textsl{Planck} Collaboration 2013). In round numbers, the parameters read:
matter density $\Omega_M = 0.3$, baryon density $\Omega_B = 0.04$, Hubble
constant $h=H_0/100$ km s$^{-1}$ Mpc$^{-1}= 0.7$, and cosmic mass variance
$\sigma_8=0.8$ on a scale of $8\, h^{-1}$ Mpc.

\section{IntraCluster Plasma in the DM Potential Wells}

The overall process of gravitational collapse that builds up the DM halos is
very relevant to the physics of the ICP, and deserves some preliminary
discussion.

\subsection{DM gravitational wells}

We recall that the formation of a DM halo starts from a small amplitude,
large-scale overdensity; this initially shares the Hubble expansion until it
detaches under the pull of its own gravity, turns around, collapses and
eventually virializes with a density contrast $\delta\rho/\rho\approx 200$
over the average background (see Peebles 1993).

The updated view of the process envisages \textit{two} stages, concurrently
found in many state-of-the-art numerical simulations and in semianalytic
studies (e.g., Zhao et al. 2003, Fakhouri et al. 2010, Wang et al. 2011, Lapi
\& Cavaliere 2011). The first stage is constituted by an early fast collapse
at $z\gtrsim 1$ of the cluster core including many major mergers; these
events reshuffle the gravitational potential, and cause the DM to undergo
(incomplete) dynamical mixing and relaxation over a wide radial range $r\sim
R/2$. After a transition redshift $z_t\sim 1$, this is followed by a long
stage of slow, inside-out development of the halo outskirts fed by diffuse
accretion and punctuated by residual mergers; these events little affect the
inner potential well, but contribute most of the overall mass.

This view of the hierarchical clustering process substantiates the schematic
lore concerning the gross behavior (with implied considerable variance) of
the mass-dependent epoch for halo formation $1+z_t\propto M^{-(n+3)/6}$; here
$n$ is the effective power spectrum index of the initial DM density
perturbations, that ranges from $-1.5$ for groups to $-1$ for rich clusters
(cf. Peebles 1993). Note that the two-stage picture is borne out also in the
galaxy domain by the violent, dust-enshrouded star formation activity at high
redshifts $z\gtrsim 1.5$ in the progenitors of massive ellipticals, as
recently pinpointed by Lapi et al. (2011) from the \textsl{Herschel}
satellite data.

After the turmoil of the initial collapse has subsided in the core, at
$z\lesssim z_t$ a quasi-static development sets in, resulting in the growth
of a roughly spherical and scale-invariant mass distribution. It is now
agreed (e.g., Navarro et al. 2010, Wang et al. 2011) that the DM density
profile providing the gravitational potential wells is accurately rendered in
terms of a S\'ersic-Einasto (see S\'ersic 1963, Einasto 1965, Lapi \&
Cavaliere 2011) shape
\begin{equation}
\rho(r)/\rho(r_{-2})=(r/r_{-2})^{-\tau}\, e^{-{2-\tau\over \eta}\,
[(r/r_{-2})^\eta-1]}~.
\end{equation}
This is written in terms of the reference radius $r_{-2}\sim $ a few $10^2$
kpc where the dependence $\rho(r)\propto r^{-2}$ locally holds, with the
parameters $\eta\approx 0.34$ describing the middle curvature of the profile
and $0<\tau<0.9$ expressing the slope of the mild inner cusp. The above shape
is now competing with the standard Navarro, Frenk \& White (1997) formula,
since it is concurrently indicated by high-resolution $N$-body simulations
(see Genel et al. 2010, Navarro et al. 2010, Wang et al. 2011), and by
increasing observational evidences from gravitational lensing (like in Coe et
al. 2012, Newman et al. 2013a and references therein). It is also
substantiated by many theoretical analyses of the halo equilibrium based on
the static Jeans equation (e.g., Taylor \& Navarro 2001, Dehnen \& McLaughlin
2005, Lapi \& Cavaliere 2009a), and by semianalytic models for the underlying
process of dynamical relaxation (e.g., Zukin \& Bertschinger 2010, Lapi \&
Cavaliere 2011).

The DM radial distribution in Eq.~(4) is represented in Fig.~1, together with
powerlaw piecewise approximations. Note that the outer slope is steep enough
to ensure a \emph{finite} total mass, while the inner cusp is mild enough to
ensure a vanishing central force and so a \emph{round} minimum in the
gravitational potential. The basin-like profile of the latter is also shown
in Fig.~1 for later use, and so is the peaked profile of the velocity
dispersion $\sigma_r^2(r)$ that is related to the `cold' nature of the DM
particles. Of course, on small scales the potential profiles may be indented
down by local baryonic contributions from bright galaxies, while a central
cusp in the total density profile may be reestablished by the central massive
galaxies; at the other end, the outer shapes will be increasingly stirred by
non-static conditions, as observed by Newman et al. (2013a,b).

The overall halo extent (given by the current virial $R$ defined in Eq.~[1])
relative to the core (given by $r_{-2}$) is marked by the so called
`concentration' parameter $c\equiv R/r_{-2}$. In fact, during the development
from $z_t$ to the observation redshift $z$ the concentration increases (see
Zhao et al. 2003, Prada et al. 2012) approximately following $c\approx 3.5\,
(1+z_t)/(1+z)$, and so provides an estimate for the effective dynamical age
of a cluster or a group.

We stress that the resulting DM halos are closely \emph{self-similar}, i.e.,
halos of different masses in the range from rich clusters to small groups are
close to rescaled versions of each other, with the scaling provided by the
depth $G\,M/R \propto M^{2/3}$ of the potential wells but with otherwise
similar profiles. The remaining differences in shape are limited: the group
halos feature slightly flatter outskirts and central cusps, but these
peculiarities are easily swamped into the effects of angular momentum and of
baryon contributions (cf. Lapi \& Cavaliere 2011); the concentration $c$ are
higher on average for the generally older groups.

Standard reference points in the DM density profile are provided by the radii
$R_\Delta$ wherein the average overdensity takes on a given value $\Delta$
relative to the critical Universe; frequently used values read
$R_{500}\approx R/2$ and $R_{200}\approx 4\,R/3$. The absolute densities
scale with the redshift $z$ following $\rho(z)/\rho_0=\Omega_M\,
(1+z)^3+\Omega_\Lambda\equiv E^2(z)$.

Note for future use that the cumulative mass $M(<r)$ can be derived by direct
integration of Eq.~(4), with explicit expressions given by Lapi \& Cavaliere
(2011).

\subsection{Baryons in the DM wells}

One may wonder whether the baryon dispositions within the DM halos also
follow a self-similar pattern, as proposed by Kaiser (1986) and discussed by
Rosati et al. (2002), Lapi et al. (2005) and many others. This would imply
the ICP to passively follow the DM gravity pull so as to feature $T\propto
\sigma_r^2$ and $n(0)\propto \rho(0)$ with invariant profiles of $n(r)$ and
$T(r)$. On the other hand, signals pointing to broken similarity have been
highlighted by Ponman and collaborators (see Osmond \& Ponman 2004, and
references therein); in fact, groups differ from clusters for emitting in X
rays quite lower core luminosities than expected from extending down to their
temperatures $k_B T < 2$ keV the self-similar cluster scaling $L_X \propto
T^2$. The latter obtains from the dependencies of the continuum emission $L_X
\propto n^2\, T^{1/2}\, R^3$ combined with the scaling $R^2 \propto T$
derived at $z \sim 0$ from Eqs.~(1) and (2), on assuming a constant baryon
fraction; if anything, $L_X$ ought to be larger and the scaling flatter upon
including the intense line emissions at group temperatures $k_B T<2$ keV (see
Cavaliere et al. 2005, their Fig. 1).

Such a deficit in group emissions correlates with an excess in the ICP
`entropy' (actually the adiabat; see Bower 1997) that is defined as
\begin{equation}
k\equiv k_B T/n^{2/3}~
\end{equation}
and is related to the standard specific entropy $s$ by $k\propto e^{2 s/3
k_B}$. In fact, the values of $k$ computed from $n$ and $T$ observed in group
cores lie considerably above the self-similar expectation for the core
entropy $k\propto T$ (see Cavaliere et al. 2005, their Fig. 2).

We stress that entropy constitutes the \emph{key} state variable for the ICP,
since it records gains of thermal energy and radiative losses into the vast
cold Universe, while being obviously insensitive to adiabatic compressions
and expansions. On a heuristic approach, entropy quantifies the resistance of
the ICP to compression or sinking into the DM potential wells (cf. Voit
2005). In an \emph{active} view of the ICP, various processes of entropy
production and radiative losses will combine and cause the ICP to react in
different ways to the gravity pull. The observed levels range around $k\sim
10 - 100$ keV cm$^2$ in group and cluster cores, and increase by factors
$5-10$ into the outskirts. Such levels correspond to specific energies of
some keVs per particle at densities $n \sim 10^{-3}$ cm$^{-3}$, which sum up
to the huge overall thermal energies around $10^{64-65}$ ergs anticipated in
\S~1. In the present context, entropy excesses in the shallow group cores
imply lower densities and so emission deficits as observed.

How do these simple considerations withstand the recent advances in the
databases and in their physical understanding? As to the latter issue, it has
been convincingly argued by Voit (2005) that simple schemes meant to tame or
explain out such large numbers are doomed to fail. For example, stellar
energies fall substantially short of sufficiently preheating the gas prior to
its infall, on account of the limited amount of star formation and SN
explosions reckoned for the intergalactic medium, or actually observed within
the clusters; the latter constraint also rules out any major, indirect
increase in entropy left over from condensing and burying the coldest gas
into stars (see Bryan 2000, Voit \& Bryan 2001). On the other hand, central
Active Galactic Nuclei (AGNs) with outputs enhanced by gas accreting from
their cluster/group environment can heat the ICP up to some keVs per
particle; this requires a reasonable $5\%$ energy coupling to the ICP, but
also implies a tuned balance of gas accretion vs. ejection (cf. Fabian 2012).

Concerning the first issue, as the databases widened so also did the scatter
in the core values $L_X$ and $k$ within groups, to the point of blurring the
quantitative signatures of broken similarity (see Sun 2012). In addition, all
ICP observables are projected along the line of sight and relate to positions
on the sky plane, whilst the scalings relate to a $3$-D radius; so the
integrated $L_X$ values considerably exceed their $3$-D counterparts, while
easily amplifying any intrinsic scatter, especially that arising in the
hectic outskirts. Robust signals and tests of broken self-similarity require
extended profiles with the leverage provided by the range, taking advantage
of the resolution of current X-ray space telescopes that attain arcsecs with
\textsl{Chandra} and \textsl{XMM-Newton}.

We conclude that understanding the ICP physics has to come to grips with
closer modeling of the thermal state and radial dispositions, based on
gravitational energy and directly keyed to an entropy spine; a relevant
unifying view will only result as a last, overarching step. We shall begin
with ICP in overall quasi-static thermal equilibrium within spherically
symmetric DM potential wells, where the pressure $p = n\,k_B T/\mu$ obeys the
equation
\begin{equation}
{1\over m_p\, n}\,{\mathrm{d}p\over \mathrm{d}r}=-{G\, M(<r)\over r^2}~,
\end{equation}
over times longer than the sound crossing $R/(5\, k_BT/3\, \mu
m_p)^{1/2}\lesssim R/\sigma\sim 1$ Gyr. Clearly, to close this equation and
solve for our primary variable $p(r)$ a local relation between $n$ and $p$ is
needed; this is provided just by the entropy $k \propto p/n^{5/3}$.

Next we discuss how entropy is produced at the center and in the outskirts of
the ICP under the gravitational energy drive during the two stages of the
halo formation processes. We begin with rich clusters in the mass range
around $10^{14-15}\, M_\odot$, and postpone to \S~4.3 the conditions
prevailing in poor clusters and groups with masses $<10^{14}\, M_\odot$. In
addition, we focus first on thermal conditions, and defer to \S~5 the
definition and use of entropy when non-thermal, turbulent contributions are
included.

\subsection{Baryons in cluster outskirts}

In a cluster-sized halo formed at $z_t\approx 1$, the outskirts are still
developing at redshifts $z\lesssim 0.5$. The development (outlined in Fig.~2)
implies a continued gravitational \emph{inflow} of DM across the virial
boundary, settling at radii $r\sim r_{-2}$; meanwhile, outer intergalactic
gas also flows into the forming cluster. At variance with the collisionless
DM, the gas inflow is mostly halted within a few proton mean-free paths
$\lambda_{\rm pp}\approx 10^2$ kpc from the virial boundary, and is
thermalized in a layer of accretion \emph{shocks} hovering there (see Lapi et
al. 2005). These constitute the main means for converting gravitational into
thermal energy, so producing much of the ICP entropy.

At a shock the net outcome from the classic Rankine-Hugoniot jump conditions
(detailed in Lapi et al. 2005, see their Appendix B) is to raise the density
by modest factors $n_2/n_1\lesssim 4$ from the intergalactic levels $n_1\sim
10^{-5}$ cm$^{-3}$, while boosting the temperature from the `field' value
$k_B T_1\sim 10^{-1}$ keV by factors that approach $T_2/T_1\approx
\mathcal{M}^2/3$ for $\mathcal{M}^2\gtrsim 3$ (see Fig.~3), here referred to
as the regime of `strong' shocks. The inflow Mach number $\mathcal{M}\equiv
v_1/c_{s}$ appearing here is the ratio of the gravitational inflow velocity
$v_1\approx (G\, M/R)^{1/2}\sim 10^3$ km s$^{-1}$ to that of sound $c_{s}\sim
10^2$ km s$^{-1}$ in the preshock medium.

As a result of Eq.~(5), the entropy of the intergalactic gas (itself
conserved during the stretches of adiabatic cosmological expansion; cf. Kolb
\& Turner 1990) is also boosted from levels $k_1\lesssim 10^2$ keV cm$^2$ to
$k_R\gtrsim 10^3$ keV cm$^2$, see Fig.~3. Past the boundary shocks, and in
the absence of other energy sources, the entropy of the accreted gas shells
will be conserved and \emph{stratified}. In other words, the radial entropy
distribution preserves the memory of the past development.

Soon after the core collapse, when the inflow is still sustained and strong
shocks \emph{efficiently} thermalize the infall energy, this process produces
an outer entropy ramp $k(r)\propto r^a$ with $a\sim 1$. In round terms, such
a slope obtains mainly because the boundary temperature $T_R$ from
gravitational heating grows under the pull of the progressively increasing
cluster mass.

In closer detail, the entropy slope $a$ has been derived by Cavaliere et al.
(2009) on matching the jumps at the boundary shocks and the adjoining
hydrostatic equilibrium of the ICP (see Eq.~[6]) to obtain
\begin{equation}
a\approx 2.5-0.5\, b_R
\end{equation}
in terms of the ratio $b_R\equiv \mu m_p\, v_R^2/k_B T_R$ of the
gravitational to the thermal energy at the boundary. This reads $b_R=3\,
v_R^2/\mathcal{M}^2\, c_{s}^2$ when `strong' shocks with Mach numbers
$\mathcal{M}^2> 3$ \emph{efficiently} thermalize the infall energy $m_p\,
v_1^2/2$ to yield the ceiling temperature $k_B T_R\approx \mu m_p\, v_1^2/3$.
\footnote{Here we have neglected the residual kinetic energy corresponding in
the shock frame to $v_2^2/v_1^2=n_1^2/n_2^2\approx 1/16$, to be taken up in
\S~5.1. When this is accounted for, one obtains $b_R\simeq 3\,
v_R^2/\mathcal{M}^2\, c_{s}^2\, [1-3/\mathcal{M}^2]$.}.

The values of $a$ (clearly smaller than $2.5$) sensitively depend on $b_R$;
to see how, it is convenient to relate the inflow Mach number
$\mathcal{M}^2\approx 2\, \Delta\phi\, (v_R^2/c_{s}^2)$ to the relevant
potential drop $\Delta\phi\equiv \Delta\Phi/v_R^2$ (normalized with the
squared circular velocity $v_R^2$) from the turnaround $R_{\rm ta}\approx 2\,
R$ to the shock radius $R_s\approx R$; this yields $b_R=3/2\Delta\phi$
depending on the outer shape of the well.

In the way of a significant example, an initial scale-invariant perturbation
$\delta M/M\propto M^{-\epsilon}$ yields the outer potential drop $\Delta
\phi\approx [1-(R_s/R_{\rm ta})^{3\epsilon-2}]/(3\epsilon-2)$, as depicted in
Fig.~2; for values $\epsilon\approx 1$ that describe the fast collapse of the
core as a whole, values $\Delta \phi\approx 1-R/R_{\rm ta}\approx 0.5$ and
$a\approx 1$ obtain. On using the detailed shapes of the gravitational
potentials associated to Eq.~(4) and represented in Fig.~1, values $\Delta
\phi\approx 0.6$ and $a\approx 1.1$ are found (see also Tozzi \& Norman 2001,
Lapi et al. 2005, Voit 2005).

However, as the cluster outskirts grow the inflows through the boundary
dwindle and slow down considerably; this occurs when the accretion feeds on
the tapering wings of a DM perturbation over the background, itself lowering
under the accelerated cosmic expansion at low $z$ (see Fig.~2). In addition,
the shocks will outgrow the virial radius and move into a region of flatter
$\Delta \phi$. These conditions are conducive to lower the inflow Mach
numbers to values $\mathcal{M}^2<3$ and so \emph{weaken} the shock jumps.
Thus the latter will produce \emph{less} entropy (see Fig.~3), with boundary
values lowered to levels $k_R\lesssim 10^3$ keV cm$^2$ and slopes
\emph{flattened} considerably below $a\approx 1$ or even bent over (Lapi et
al. 2010, Cavaliere et al. 2011a).

To make our discussion explicit, we pursue the above example based on the DM
perturbation $\delta M/M\propto M^{-\epsilon}$; according to this, outskirts
development corresponds to effective values of $\epsilon$ growing above $1$,
as is seen on considering the accretion rate $\dot M$. A shell $\delta M$
enclosing the mass $M$ will collapse when $\delta M/M$ attains the critical
threshold $1.686\, D(t)$ in terms of the linear growth factor $D(t)$, cf.
Weinberg (2008). Accordingly, the shape parameter $\epsilon$ also governs the
mass buildup after $M\propto D^{1/\epsilon}(t)$; on the other hand, the
growth factor may be represented as $D(t)\propto t^d$, with $d$ ranging from
$2/3$ for $z\gtrsim 1$ to approach $1/2$ as $z$ lowers to $0.2$ and below. So
the outskirts develop from the inside-out, with accretion rates $\dot
M/M\approx d/\epsilon t$ that decrease for $\epsilon$ exceeding $1$ as the
accretion involves the perturbation wings, and as $d$ decreases toward $1/2$
at late cosmic times in the accelerating Universe. In such conditions, the
effective potential drop $\Delta \phi\approx [1-(R_s/R_{\rm
ta})^{3\epsilon-2}]/(3\epsilon-2)$ quoted above lowers so does the infall
speed $v_1^2\propto \dot M^{2/3}\, (\Delta\phi)^{2/3}$ proportional to
$d^{2/3}/\epsilon^2$, including the effects of shock outgrowth beyond $R$ by
decreasing ram pressure, see Fig.~2.

Thus the outer entropy profiles will \emph{flatten out} on the timescale set
by the halo development, as the concentration grows to values $c\gtrsim 6$
from the initial $c\approx 3.5$ set soon after the core collapse at $z_t$;
the time involved will amount to some $5$ Gyrs for a cluster collapsed at
$z_t\approx 1$ and observed at $z\approx 0.15$. On the other hand, generally
\emph{flatter} entropy slopes $a\approx 0.6-0.8$ apply whenever the boundary
shocks are weaker due to less supersonic inflows. This condition may occur in
clusters, when the infalling gas is preheated as it runs down filaments of
the Large Scale Structure (e.g., Valageas \& Silk 1999, Wu et al. 2000,
Scannapieco \& Oh 2004, McCarthy et al. 2008). It often applies to groups
where lower infall is driven by a smaller mass with a generally shallower and
flatter potential well relative to clusters (see Fig.~2 in Lapi \& Cavaliere
2009a; also Sun 2012). Observational outcomes will be discussed in \S~4.3.

\subsection{Baryons in cluster cores}

At the other end, the central entropy originated in the early fast collapse
is set at $k_c\sim 10^2$ keV cm$^2$, not far above the intergalactic levels.
This is because during the initial fast collapse the temperatures in the
virialized core are raised to $k_B\,T\approx G\,m_p\, M(<r)/10\,r \sim$ a few
keVs weakly depending on mass, while the ICP is thickened to some $n\sim
10^{-3}$ cm$^{-3}$, in step with the overdensities $\delta\rho/\rho \gtrsim
2\times 10^2$ marking all virializing structures. These levels of $k_c\sim
10^2$ keV cm$^2$ are similar from clusters to groups, so emulating the
outcome from a general preheating, which would require very large, diffuse
energy inputs into the intergalactic gas.

But in clusters such initial entropy levels at the center may be subsequently
\emph{eroded} or even erased due to the cooling by the observed
bremsstrahlung radiation and line emissions for $k_B T\lesssim 2$ keV. The
cooling timescale for a single-phase ICP at $k_B T\gtrsim 2$ keV (cf. Sarazin
1988) reads $t_{\rm cool}\approx$ $30\, (k_B T/{\rm keV})^{1/2}$
$(n/10^{-3}~{\rm cm}^{-3})^{-1}$ Gyr. While in the low-density outskirts
radiative cooling is slow and little relevant, it is speeded up in the dense
central ICP, so that in some $5$ Gyr the initial levels $k_c$ may be
considerably lowered down to $\sim 10$ keV cm$^2$. Whence cooling would
become so fast as to match the dynamical times $\sim 10^{-1}$ Gyr, to the
effect of impairing the thermal pressure support; the process is even faster
in multi-phase ICP with a considerable cold component as stressed by Rossetti
\& Molendi (2010).

This leads to ICP condensation and to cooling faster yet, so as to start an
accelerated settling to the cluster center and onto the central galaxies (the
classic `cooling catastrophe'; e.g., White \& Rees 1978, Fabian et al. 1984,
Blanchard et al. 1992), were it not for renewed energy injections (as widely
entertained by, e.g., Binney \& Tabor 1995, Cavaliere et al. 2002, Lapi et
al. 2003, Voit \& Donahue 2005, Tucker et al. 2007, Hudson et al. 2010). Such
injections occur when the condensing ICP reaches down into the galactic
nuclei and onto their central supermassive black holes, to trigger or feed a
loop of intermittent starbursts and AGN activities. In the form of gentle
bubbling or moderate outbursts over some $10^{-1}$ Gyr, these can stabilize
the time-integrated values $k_c$ at levels around $10$ keV cm$^2$ (see, among
others, Roychowdhury et al. 2004, Ruszkowski et al. 2004, Sijacki \& Springel
2006, Ciotti \& Ostriker 2007, McNamara \& Nulsen 2007, Fabian 2012).

In addition, the levels of $k_c$ may be abruptly \emph{raised} up to some
$10^2$ keV cm$^2$ when substantial energy injections $\Delta E$ occur into
the ICP from violent outbursts of AGNs in central galaxies, and even more
from deep mergers. These injections launch through the central ICP outgoing
blastwaves bounded by a leading shock with Mach number given by
$\mathcal{M}^2 \approx 1 + \Delta E/E$ in terms of the central ICP thermal
energy $E\approx 2\times 10^{61}\, (k_BT/\mathrm{keV})^{5/2}$ erg (see Lapi
et al. 2005, their Fig.~7). Strong outgoing shocks with $\mathcal{M}^2\gtrsim
3$ require injections $\Delta E\gtrsim 2\, E$, i.e., some $10$ keV per
particle. This may be the case for deep major mergers, more easily than for
AGNs powered by a supermassive black hole of $5\times 10^{9}\, M_{\odot}$
with just some $5\%$ of the discharged energy effectively coupled to the ICP,
as argued by Lapi et al. (2005).

Blasts that preserve overall virial equilibrium may still leave a
long-lasting imprint onto the central ICP in the form of an entropy
\emph{hot} spot spread out to a radius $r_f\sim 10^2$ kpc, where the blast
has expanded, stalled and degraded into sound waves (see McNamara \& Nulsen
2007, Fabian et al. 2011; also Fusco-Femiano et al. 2009). Even stronger if
rarer energy injections with $\Delta E\gg E$ will be produced when major,
head-on mergers (e.g., McCarthy et al. 2007, Norman 2011) deposit at the
center large energies around some $10$ keV per particle, leading to entropy
levels up to $\sim 5\times 10^2$ keV cm$^2$.

\section{Hydrostatic Equilibria of the ICP}

The above processes for entropy production and stratification combine into
the basic pattern proposed by Lapi et al. (2005) and Voit (2005)
\begin{equation}
k(r)=k_c + k_R\, (r/R)^a~.
\end{equation}
This rises from the central `floor' $k_c\sim 10^{1-2}$ keV cm$^2$ into a ramp
with slope $a\lesssim 1$ toward the boundary value $k_R\sim 10^3$ keV cm$^2$,
and is outlined in Fig.~4.

\subsection{The entropy-based Supermodel}

Entropy constitutes not only the thermodynamical handle to the behavior of
the ICP, but also the operational key to `close' Eq.~(6) of hydrostatic
equilibrium on expressing the density as $n(r)\propto [p(r)/k(r)]^{3/5}$. The
resulting first-order differential equation for the primary variable $p(r)$
is linear, and straightforwardly integrates (e.g., Dwight 1961) to yield
\begin{equation}
{p(r)\over p_R}=\left[1+{2\, G\, m_p\over 5\, p_R^{2/5}}\,\int_r^R{\mathrm
d}x~{M(<x)\over x^2\, k^{3/5}(x)}~\right]^{5/2}~,
\end{equation}
as proposed by Cavaliere et al. (2009), and used by Allison et al. (2011).

Eq.~(9) has the stand of a \emph{theorem} in hydrostatics (with the attendant
asymptotic corollaries spelled out below), once the basic pattern of $k(r)$
is pinpointed; as to the latter, a physical \emph{model} is provided by the
two-stage halo formation discussed in \S~2.1, and is presented in Eq.~(8). In
the following, this approach is tested on \emph{observables}. To begin with,
the run of $p(r)$ integrated along the l.o.s. is directly probed with the
Sunyaev-Zel'dovich (1980; SZ) effect, to be discussed in \S~4.5. On the other hand,
from $p(r)$ the model profiles of the density $n(r)\propto [p(r)/k(r)]^{3/5}$
and of the temperature $T(r)\propto p^{2/5}(r)\, k^{3/5}(r)$ obtain in closed
forms, \emph{linked} together through the same underlying $k(r)$; below these
observables will be compared first with the X-ray data. Meanwhile, note that
among these variables $p$ has the weakest dependence on $k$, with
implications taken up in \S~4.4 and 5.

Some simple asymptotics is reported here for later use. At the cluster
center, the integral term in Eq.~(9) behaves like $k_c^{-0.25}$ as discussed
by Cavaliere et al. (2009), to imply the scaling laws $p_c\propto
k_c^{-3/5}$, $T_c\propto k_c^{0.35}$ and the projected X-ray brightness
$S_X\propto n_c^2\, T_c^{1/2}\propto k_c^{-1.8}$. These show how for
\emph{low} values of $k_c\lesssim 40$ keV cm$^2$ the central temperature
\emph{drops} to a non-zero value $T_c$ before rising to a middle maximum
marked by $\mathrm{d}T/\mathrm{d}r=0$ at $r\sim 0.1\,R$; meanwhile, the
central emissivity \emph{peaks} to a finite value due to the finite ICP
pressure. Such features mark the standard, relaxed cool-core (CC) clusters as
defined by Molendi \& Pizzolato (2001; see also Hudson et al. 2010). On the
other hand, \emph{high} values of $k_c$ imply \emph{flat} emissivity profiles
together with a central temperature \emph{plateau} or high \emph{rise},
typical of the many unrelaxed non-cool-core (NCC) clusters. We illustrate
these model morphologies in Figs.~5 and 6. We add that the central cooling
time in a single-phase ICP equilibrium may be expressed in terms of the
entropy level $k_c$ only, to read
\begin{equation}
t_c \approx 0.5\, (k_c/15\, \mathrm{keV~cm^2})^{1.2}~\mathrm{Gyr}~;
\end{equation}
this implies that high levels of $k_c\gtrsim 10^2$ keV cm$^2$ require long
timescales $\gtrsim 5$ Gyr to be eroded.

In the outskirts, instead, Eq.~(9) yields the scaling laws $p(r)\propto
r^{2\,a-5}$ and $T(r)\propto r^{7\, a/5-2}$; these show that \emph{flatter}
entropy slopes imply \emph{steeper} declines of the pressure and of the
temperature.

To obtain the full \emph{profiles}, one inserts the entropy pattern of
Eq.~(8) into Eq.~(9) and the related expressions for $n(r)$ and $T(r)$ given
above. These relations constitute what Cavaliere et al. (2009) dubbed
`Supermodel' (SM) in a warm mood prompted by its ability to include as
particular instances several previous models, and to describe both the CC and
the NCC configurations of the ICP in terms of a few physical parameters. The
latter enter the basic entropy run in Eq.~(8) and read: the floor $k_c$ in
the core, and the Mach number $\mathcal{M}^2$ of the boundary shocks
governing the ramp $a$ after Eq.~(7). They are pinned down from fitting the
data that concern the X-ray brightness $S_X$ and temperature $T(r)$ as
discussed next, or concern directly the pressure $p(r)$ from observations of
the SZ effect discussed in \S~4.5. In all such cases the boundary values
$T_R$ and $p_R$ are simply related to $\mathcal{M}^2$ (just proportional for
$\mathcal{M}^2\gtrsim 3$) by the classic jump conditions at the outer shocks
recalled in Fig.~3 and its caption.

\subsection{Relationships to simple models}

The above description of the hydrostatic equilibrium after Eq.~(9) may be
related to standard, simpler models where the entropy is just assumed to be a
functional of the density alone. Specifically, the simplest model corresponds
to an isothermal equation of state $k\propto n^{-2/3}$, which yields
$n\propto e^{\beta\,\Delta\phi}$ in terms of the gravitational potential drop
$\Delta\phi\equiv [\Phi(R)-\Phi(r)]/\sigma_r^2$ normalized to $\sigma_r^2$,
and of the ratio $\beta\equiv \mu m_p\, \sigma_r^2/k_B T\approx 0.7$ between
the DM and the ICP scale heights. If $\sigma_r(r)$ were constant, i.e., the
DM were itself `isothermal', then the simple expression $n(r)\propto
\rho^\beta(r)$ would apply (Cavaliere \& Fusco-Femiano 1976). Such an
`isothermal $\beta$-model' (widely taken up since Jones \& Forman 1984) works
well for the central and middle regions of NCC clusters like Coma, which
indeed are roughly isothermal on a scale of several $10^2$ kpc (e.g.,
Churazov et al. 2012; see also our Fig.~5).

On the other hand, one may try a standard polytropic assumption $k\propto
n^{\Gamma-5/3}$ in terms of a constant macroscopic adiabatic index
$1\leq\Gamma\leq 5/3$ with the bounds corresponding to the isothermal and to
the convectively mixed conditions, but with little specific physics in
between. When inserted in the equilibrium this relation provides the solution
$n\propto T^{1/(\Gamma-1)}\propto [1+{\Gamma-1\over \Gamma}\,
\beta\,\Delta\phi]^{1/(\Gamma-1)}$, see discussion by Cavaliere \&
Fusco-Femiano (1978); Lea et al. (1973) and Gull \& Northover (1975) had
argued for the specific values $\Gamma\approx 1.35$ and $5/3$, respectively.
It is now clear that the model with $\Gamma\approx 1.2$ works reasonably in
the body of massive CC, relaxed clusters; on the other hand, to represent
their inner behavior values $\Gamma< 1$ would be required since $T\propto
n^{\Gamma-1}$ must be on its rise toward its middle maximum while $n(r)$ is
still lingering about its own central maximal value.

But then a closer description of CC clusters is provided by the `mirror
dispersion' approach (proposed by Cavaliere \& Fusco-Femiano 1981 and tested
by Hansen \& Piffaretti 2007); this envisages the temperature profile
$T(r)\propto \sigma^2(r)$ to mirror the peaked behavior of the DM velocity
dispersion $\sigma(r)$. The net outcome is $n(r)\propto \rho^{\beta}(r)\,
\sigma^{2\,(\beta-1)}(r)$, which in a polytropic-like interpretation
corresponds to $\Gamma< 1$ in the central range. A specific instance is
illustrated in the top panel of Fig.~6.

These piecewise representations are actually incorporated and unified by the
SM; in fact, the latter may be represented in terms of an effective, radially
varying polytropic index $\Gamma(r)=5/3+{\rm d}\log k/{\rm d}\log n$ ranging
from values $\approx 0.5$ at the center to $\approx 1.2$ into the outskirts
(see Fig.~3, bottom panel, in Cavaliere et al. 2009). Thus as a fitting tool
the SM performs \emph{uniformly} better throughout the full range from CC to
NCC clusters than any of those simpler models, yet with its fewer, intrinsic
parameters $k_c$ and $a$. Moreover, as the SM is based on the physical
entropy pattern of Eq.~(8) linked to the DM halo development (see \S~3.1),
the derived observables shed light on these dynamical processes relevant to
the ICP thermal state, and suggest the physically based cluster
classification that will be proposed in \S~4.

\subsection{Data fitting, and beyond}

With the full SM, one uses Eq.~(9) including the basic entropy pattern
Eq.~(8) with the two free parameters $k_c$, $a$ (plus the related boundary
value $k_R$ closely proportional to $\mathcal{M}^2$), and derives the radial
profiles of density and temperature normalized at the boundary; a very fast
algorithm managing this task is available at the URL
\texttt{http://people.sissa.it/ $\sim$lapi/Supermodel/}. Thus one can perform
fits to the projected, emission-weighted temperature and/or brightness data
(including instrumental bandpass), test them with the use of a standard
biparametric procedure for $\chi^2$ minimization (e.g., \textsl{MPFIT} by
Markwardt 2009), and derive the bestfit values of the two basic entropy
parameters with their uncertainty ranges.

Such fits can be performed over the full radial range covered by the current
X-ray data. In a number of clusters observed with the low-background
instrumentation of \textsl{Suzaku} (e.g., A1795 and A1835), the X-ray data
extend toward the virial radius $R$, though with the systematics debated by
Eckert et al. (2011, 2013) and Walker et al. (2012, 2013). In other instances
observed with \textsl{XMM-Newton} the data are limited to around $r\sim R/2$.
Note that the shape parameters $k_c$ and $a$ may be obtained from fitting
either the temperature or the brightness profile, and the results turn out to
be consistent within the respective uncertainties. Ordinarily the brightness
data are substantially more precise, and allow a more robust reconstruction
of the entropy profile.

The outer scale $R$ is usually provided by independent observations such as
red-sequence termination or gravitational lensing, and so is $c$ (e.g.,
Medezinski et al. 2007, Broadhurst et al. 2008). On the other hand, fits to
the X-ray brightness can also determine the DM concentration $c=R/r_{-2}$
that enters the SM formalism through $M(<r)$, so as to provide a handle from
the ICP observables to the dynamical age and history of the host DM halos.
Such determinations of $c$ are mainly based on outer brightness data (see
Fig.~7), so they are independent of the inner entropy level $k_c$; they are
fast yet robust. The results turn out to be consistent with direct but
laborious measurements based on gravitational lensing (cf. Broadhurst et al.
2005, Lemze et al. 2009, Lapi \& Cavaliere 2009b), yet are less biased than
the latter by the prolateness effects discussed, e.g., by Corless et al.
(2009).

Thus the basic parameters entering the entropy pattern are calibrated from
fitting with the SM the projected observables (brightness and
emission-weighted temperature) directly computed from $3$-D profiles of
$n(r)$ and $T(r)$, with no need for delicate data deprojections (widely
discussed by Kriss et al. 1983, Yoshikawa \& Suto 1999, Cavaliere et al.
2005, Croston et al. 2006, Urban et al. 2011). Note that throughout most of
the cluster volume these results are robust against reasonable deviations
from spherical symmetry, hydrostatic equilibrium, and strictly smooth
accretion. In fact, in the inner regions any geometrical asymmetries like the
merger-related ones, are smoothed out on a crossing timescale, shorter than
the time taken by cooling to erase entropy excesses of some $10^2$ keV
cm$^2$. In the middle regions, approximately spherical symmetry of the ICP is
indicated by various simulations (e.g., Lau et al. 2011). In the outer
regions, the accretion is often dominated by minor mergers or truly diffuse
matter funneled by filaments, as shown in detail by the simulations of Wang
et al. (2011, see their Fig.~7) and discussed in \S~4.3.

\section{Physical Outcomes from SM Analyses}

We have just seen how the parameters specifying the entropy distribution can
be derived from fitting the X-ray data with the SM. The results from the
analysis on $12$ clusters with high-quality X-ray data (that implies
$z\lesssim 0.2$) are collected in Table~1; specific examples of the fits with
the SM are illustrated in Figs.~5, 6, and 7. These clusters are apparently
parted into two main blocks on the basis of their $k_c$ values running from a
few $10^1$ to a few $10^2$ keV cm$^2$; within each block, the members are
ordered on the basis of their $a$ values. These two blocks turn out to be
also parted in terms of their DM concentration $c$. Such an ordering points
toward \emph{correlations} between these basic physical parameters, to be
discussed next.

\subsection{Correlations}

In the top panel of Fig.~8 we illustrate the central levels $k_c$ vs. the
outer slopes $a$ taken from Table~1; we find values $a\gtrsim 1$ for NCCs
(red dots), and appreciably lower ones for CCs (blue dots). It is seen that
$a$ correlates on average with $k_c$; statistical tests detailed in Fig.~8
and its caption show that chance occurrence of such a correlation is limited
to under $9\%$ probability, while that of `outliers' (objects with $k_c\geq
30$ keV cm$^2$ and $a\leq 0.6$) is around $5\%$ on average.

In the bottom panel of Fig.~8 we illustrate the values of the outer slope $a$
vs. the concentration $c$, as taken from Table~1. Low values of $a$
correspond to high values of $c$, marking a long lifetime from the formation
$z_t$ to the observation redshift $z\approx 0$, see \S~2; such an
anti-correlation between $a$ and $c$ turns out to be even more statistically
significant.

\subsection{Classes, toward a Grand Design}

The above results indicate that many rich clusters like those listed in
Table~1 can be parted into two main classes, defined on the basis of high
entropy (HE) or low entropy (LE) prevailing both in the inner region and
throughout the ICP.

$\bullet$ \emph{LE clusters} feature \emph{low} entropy throughout the ICP;
this includes both a \emph{low} central baseline $k_c<30$ keV cm$^2$ and a
moderate outer level $k_R\lesssim 10^3$ keV cm$^2$, consistent with a ramp
flattening toward $a(r)<1$ outwards of $r_b/R\gtrsim 0.3$. The outcome is a
low central value of $T$ and a \emph{peak} of $T(r)$ at $r/R\lesssim 0.2$
followed by a decline outwards, particularly effective at low $z$ (e.g.,
A1795). We stress that such a class definition includes not only a central
CC state as in the standard designation, but also an associated low level
of outer entropy production. The association: low $k_c$ $-$ shallow $a$ is
to be traced back to a long lifetime of the containing DM halos, marked by
\emph{high} values of the concentrations $c\gtrsim 6$. Such a late stage in
the outskirts development is associated to dwindling inflows that produce
weaker boundary shocks with $\mathcal{M}^2\lesssim 3$ and related lower
entropy production, as discussed by Lapi et al. (2010).

$\bullet$ \emph{HE clusters} feature \emph{high} entropy throughout the
ICP; that is to say, they feature not only a high central floor $k_c\approx
3\times 10^2$ keV cm$^2$, but also an high outer level $k_R\approx
3-5\times 10^3$ keV cm$^2$, corresponding to a steep entropy ramp with
$a\gtrsim 1$ toward the outskirts. The high values of $k_c$ yield a
monotonic temperature profile $T(r)$ throughout, declining from the central
high rise or plateau into the outskirts, before a final drop toward the
boundary. This class definition includes not only a central NCC state as in
the designation introduced by Molendi \& Pizzolato (2001) and pursued by
Leccardi et al. (2010), but also an associated high level of outer entropy
production. The association arises because the \emph{young} age of the
containing DM halos, marked by low values of the concentrations $c<5$,
implies a lifetime (cf. \S~2.1) too short for the high central entropy to
be erased away and for general entropy flattening to be effective in the
outskirts.

The low $k_c$ levels proper to LEs are related to, and in fact driven by
cooling timescales $t_c$ shorter than the halo dynamical age indicated by
$c$, see Eq.~(10). In fact, the transition between LEs and HEs occurs around
$k_c\approx 40$ keV cm$^2$ corresponding to cooling times $t_c\approx $ a few
Gyrs; thereafter, fast cooling leads to an accelerated progress toward $k_c$
levels lower yet. Eventually, the levels of $k_c$ are likely to be
stabilized, on a time average basis, by the two additional physical processes
anticipated \S~2.4, i.e., intermittent AGN activity and impacts of deep major
mergers. These two modes are suggested by the broad, possibly double-peaked
distribution for the number of clusters with a given $k_c$, as observed by
Cavagnolo et al. (2009) and Pratt et al. (2010), and discussed by Cavaliere
et al. (2009).

The relationship between the classes is depicted in the \emph{evolutionary}
chart of Fig.~9, that represents the cluster Grand Design proposed by
Cavaliere et al. (2011a). This envisages clusters mainly born in an
\emph{unrelaxed} HE state of high entropy, dominated by the fast violent
collapse of the halo bulk with related strong inflows and shocks in the
infalling gas. Subsequently, on a timescale of several Gyrs they develop an
outer halo while they progress toward a \emph{relaxed} LE state; the central
entropy is lowered by radiative cooling, while the outer entropy ramp
flattens down or even bends over due to the weakened shocks and tapering
entropy production. In some cases the sequence may be halted after a few Gyrs
and \emph{reversed} by late deep mergers which \emph{rejuvenate} the central
ICP into a higher entropy state; a detailed discussion is given by
Fusco-Femiano et al. (2009).

\subsection{Predictions}

Specific predictions from our Grand Design are as follows.

$\bullet$ HE clusters are expected to feature a still incompletely developed
halo spanning only a limited radial range from the core. This implies a steep
brightness profile $S_X(r)$, with a step-wise shape of the temperature
profile $T(r)$, from high central values to a drop toward the shock layer
(see Lapi et al. 2005). We expect such conditions to be particularly sharp at
\emph{high} $z$.

$\bullet$ LE clusters at \emph{low} $z$ are expected to feature in their
entropy profiles particularly low values of $k_c$, and outer ramps flattening
down or even bending over. These will produce declining $T(r)$ profiles
outward of the middle peak following $T(r)\propto n^{2/3}(r)\, k(r)$ as
argued by Lapi et al. (2010) and supported by the \textsl{Suzaku}
observations of A1795 (cf. bottom panel of Fig.~6); a similar case may be
constituted by A2142, cf. Akamatsu et al. (2011). In such structured cases,
clearly the SM requires two more parameters in the entropy profiles as
proposed by Cavaliere et al. (2011a) and borne up by Walker et al. (2012),
namely: the position where bending sets in, and the outer entropy slope
(related to low virial value $k_R$). The outcomes are illustrated in Fig.~6,
bottom panel. These expectations have been recently borne out by a detailed
analysis of the outer entropy profiles in a sample of $11$ LE clusters, as
illustrated in Fig.~10 (from Walker et al. 2012; see also Hoshino et al.
2010, Sato et al. 2012, Ichikawa et al. 2013).

$\bullet$ From our Grand Desing we expect the evolution from HEs to LEs to
imply a \emph{lower} fraction of LEs at \emph{higher} $z$, in accord with the
evidence from Santos et al. (2010) to Sayers et al. (2013). Note that
high-$z$ LEs are more conspicuous in a X-ray analysis following up an SZ
survey, yet their observed fraction is still modest (see Santos et al. 2012,
Semler et al. 2012).

Beyond object-to-object variance, specific circumstances (limited in our
Table~1 to some $15\%$) may blur the simple bimodal classification as
proposed in \S~4.2, and alter the straightforward evolutionary path proposed
here. First, the very definition of CC clusters is recently getting
articulated into strong, intermediate, and weak CCs (see Hudson et al. 2010,
Sun 2012). Second, we recall from \S~4.2 that HEs especially at $z\gtrsim
0.5$ will feature halos and ICP in brisk and often clumpy growth; cold clumps
may easily pierce the virial shocks (similar events have been discussed by
Dekel \& Birnboim 2008 in a galactic context), and thermalize only near the
center after some sloshing across (e.g., ZuHone et al. 2010), resulting in a
enhanced inward increase of $T(r)$. Third, in clusters at low $z\lesssim 0.5$
the actual accretion rates will be particularly sensitive to the richness of
the cluster environment including its filamentary structures. So azimuthal
sectors facing filaments may retain HE-like features; on the other hand,
those facing voids develop sharper LE-like features. In particular, within
the former we expect a marked entropy \emph{flattening} for $r\gtrsim
R_{500}$ due to accretion drying out; meanwhile, in the latter sectors we
expect the entropy to be still sustained out to the layer of strong shocks
standing at the boundary. Note, in addition, that ongoing mergers easily
occur in HEs, along with possible bow shock fronts. This is our reading of
the complex features in the data presented by Simionescu et al. (2013) for
A1656, and by Kawaharada et al. (2010) and Ichikawa et al. (2013) for several
LE-like clusters.

As to groups compared with clusters, our framework leads us to envisage a
structure more HE-like in the center, and more LE-like in the outskirts. This
is because, as anticipated in \S~2.3, less supersonic inflows are driven
across the boundary by the smaller masses, and produce less entropy with
flatter slopes. Meanwhile, the shallower central wells host entropy levels
still close to some $10$ keV cm$^2$ produced by the first DM infall and then
increased by AGN outbursts; these levels are high enough to appreciably lower
the central densities and considerably decrease the emission so preventing
fast cooling. The recent data collated by Sun (2012) bear out this picture.

\subsection{Checking the Grand Design on X-ray data}

What is the current evidence of the evolutionary trend envisaged by our Grand
Design? Toward an answer, it is useful to compare in Fig.~11 the pressure
profiles derived from our SM with X-ray observations and with numerical
simulations.

The yellow shaded area illustrates the region covered by the low redshift
($z\lesssim 0.2$) clusters of the \textsl{REXCESS} X-ray sample analyzed by
Arnaud et al. (2010); the dotted blue and red lines refer to the average
pressure profiles for the subsamples of CC (relaxed) and NCC (typically
disturbed) clusters. The cyan shaded area illustrates the region covered by
hydrodynamical simulations of relaxed clusters (Borgani et al. 2004, Nagai et
al. 2007, Piffaretti \& Valdarnini 2008, Battaglia et al. 2011). Following
the standard convention, in the plot the radial scale is normalized to
$R_{500}$, while the pressure is normalized to the value $p_{500}\approx
1.8\, E^{8/3}(z)$ $(M_{500}/5\times 10^{14}\, M_{\odot})^{2/3}$ eV cm$^{-3}$,
in terms of the $z$-dependence $E(z)$ defined at the end of \S~2.1 (see also
Ettori et al. 2004).

The dashed line represents the so-called `universal' pressure profile
$p(r)/p_{500}$ $= p_0 \, (c_{500}\,x)^{-\gamma}\,
[1+(c_{500}\,x)^{\alpha}]^{-(\beta-\gamma)/\alpha}$ in terms of the variable
$x\equiv r/R_{500}$ and of the free parameters [$p_0$, $c_{500}$, $\alpha$,
$\beta$, $\gamma$]. This empirical formula had been originally proposed by
Nagai et al. (2007) to interpret the outcomes of hydrodynamical simulations
of relaxed clusters (see also Battaglia et al. 2011); subsequently, it has
been exploited by Arnaud et al. (2010) as a fitting tool to the pressure
profiles from deprojected X-ray data. With the particular parameter set [8.4,
1.177, 0.3081, 1.0510, 5.4905], it has been used to render the average
pressure profile from the real data within $R_{500}$, and the simulated ones
beyond that radius.

However, the wide variance in these X-ray data for $r\lesssim R_{500}$ shows
that such an average profile yields only an incomplete description. In fact,
the partial averages over the CC and NCC subsamples deviate upward and
downward by amounts substantially exceeding their internal variance; thus a
\emph{bimodal} description constitutes both a closer and a more effective
representation.

This is provided by the SM templates for the pressure profiles given by our
Eq.~(9) for typical HEs (entropy parameters $k_c=100$ keV cm$^2$,
$k_R=3\times 10^3$ keV cm$^2$ and $a=1.1$) and LEs ($k_c=10$ keV cm$^2$,
$k_R=10^3$ keV cm$^2$, and $a=0.7$) clusters; they are illustrated with the
same normalization in Fig.~11 by the red and blue solid lines. In the core
these feature the NCC and CC behaviors, while for $r\gtrsim R_{500}$ where
still scarce X-ray data are available the LE template agrees well with the
results from hydro$-$simulations of relaxed clusters. In the way of a
\emph{prediction}, we expect for HEs appreciably higher pressure profiles
relative to LEs, from $r\approx R_{500}$ out to the still developing boundary
at $r\approx R$.

We stress that the weak dependence of $p(r)$ on $k(r)$ noted in \S~3.1 has
two sides: on the one hand, it is bound to yield the \emph{large} difference
in entropy levels between typical HE and LE clusters; on the other hand, it
brings about a closely universal shape within each class. Additional reasons
why these two classes of physical pressure profiles can attain universality
are as follows. First, the SM profiles provided by Eq.~(9) concern primarily
$p(r)$ and depend only on the spherically \emph{averaged} entropy; as such,
they meet the requirements set by Khedekar et al. (2013) for a `bulk profile'
\emph{filtering out} possible kinds of fluctuations (cf. Zhang et al. 2009).
For example, isobaric fluctuations do not affect the large-scale $p(r)$, and
adiabatic ones do not matter in the SM. As to fluctuations in entropy, they
are physically related to cold gas advected by mergers, or to ICP heated up
by merger- or AGN-driven shocks; these features may be regarded as cold or
hot \emph{spots}, and are focused by the data of Rossetti \& Molendi (2010).
Second, we note that a minimal \emph{threshold} for these spots to imprint
the profiles $p(r)$ is set by the uncertainties $\Delta k_c/k_c$ and $\Delta
a/a$ related to the parent data fits discussed in \S~3.3; these will
certainly narrow down along the data progress, but currently are at levels
around $25\%$.

In fact, the impacts of such cold/hot spots on the spherical equilibrium can
be easily assessed from the scaling laws given in \S~3.1. In the central
regions where $p\propto k_c^{-3/5}$ applies, we expect $\Delta p/p\approx
-0.6\, \Delta k_c/k_c$ to hold; at the other end, in the outskirts where
$p\propto r^{2\, a-5}$ and $k\propto r^a$ apply, we expect $\Delta p/p\sim
-2\, \Delta k/k$ to prevail. In evaluating the net outcomes, however, it
should be kept in mind that at the center the entropy levels may differ by a
factor up to $10$, a value that marks out the cool cores of LE from the hot
ones of HE clusters; note in Fig.~11 that their central pressures differ in
fact by a factor around $5$. Indeed, the high pressures in the cool cores of
LEs may be described in terms of a giant cold spot maintained by radiative
cooling in gross balance with the energy injections from minor mergers and
large AGN outbursts. At the virial boundary, on the other hand, the entropies
may differ by factors around $3$ and the pressures by factors around $5$.
These values may be so used to infer the impact of the upward entropy jumps
caused by weak, merger-driven shocks with Mach numbers $\mathcal{M}\lesssim
2$ (e.g., Korngut et al. 2011, Ade et al. 2013, Akamatsu et al. 2012), which
amount to factors $\lesssim 1.5$ as seen from Fig.~3. The imprints of such
limited and localized hot spots are being pursued in narrow cluster sectors
analyzed at high resolutions, as discussed below.

\subsection{Testing on SZ data}

Another observable to \emph{directly} probe pressure profiles and local
enhancements is constituted by the thermal Sunyaev-Zel'dovich effect (1972,
1980; SZ). This occurs as CMB photons are inverse Compton scattered by the
hot ICP electrons, and change the radiation temperature $T_{\rm cmb}\approx
2.73$ K by an amount $\Delta T=g_\nu\, y\, T_{\rm cmb}\sim -0.5$ mK; at low
$\mu$wave frequencies a cold shadow is cast by the hot ICP electrons onto the
CMB sky. The signal is small, but intrinsically $z-$\emph{independent}
(differently from surface brightness), and is measured robustly with highly
sensitive space instrumentation and with angular resolutions now approaching
$10''$ from ground radiotelescopes.

In detail, the SZ strength is given by the Comptonization parameter
\begin{equation}
y\equiv (\sigma_T/m_e c^2)\, \int{\mathrm{d}\ell}\, p_e(r)
\end{equation}
integrated along the line-of-sight $\ell$. The spectral factor $g_\nu$
approaches the value $-2$ at low frequencies; its positive signature for
$\nu>217$ GHz, with values still significant our to a few THz, offers a
powerful cross-check on the SZ nature of the signals. The Comptonization
parameter constitutes a \emph{linear}, intrinsically $z$-independent probe of
the thermal electron pressure; the latter reads $p_e\approx p\, (2+2 X)/(3+5
X)\geq 0.5\, p$ in terms of the ICP pressure $p$, and takes on values
$p_e/p\approx 0.52$ with the cosmic hydrogen abundance $X\approx 0.76$.

The SZ observations have already probed the radial pressure profiles in
nearby individual clusters, and in more distant stacked samples (South Pole
Telescope [\textsl{SPT}] collaboration, Plagge et al. 2010; \textsl{WMAP}
collaboration, Komatsu et al. 2011; \textsl{Planck} collaboration, Aghanim et
al. 2011a, 2011b). They are also addressing the cluster contribution to the
CMB power spectrum at multipoles $\ell\gtrsim 2000$ (see Lueker et al. 2010,
Dunkley et al. 2011, Reichardt et al. 2011).

In fact, in Fig.~11 we carry on to higher $z$ our discussion in \S~4.4, on
using stacked SZ observations of redshifts $0.2\lesssim z\lesssim 0.4$
clusters (see Lapi et al. 2012); the pressure profiles from the \textsl{SPT}
stacked data (cf. Plagge et al. 2010) are represented with the green squares.
Although the uncertainties are still considerable in the outskirts, a
departure from the empirical `universal' profile stands out, and the
$z-$dependent trend toward an HE-like template clearly emerges, giving
support to the picture envisaged by the Grand Design. The same trend emerges
from the analysis of a stacked cluster sample observed with $\textsl{WMAP}$
(Komatsu et al. 2011). A similar trend is also suggested by the sample of
clusters detected with the $\textsl{Planck}$ satellite for redshift
$0.3\lesssim z\lesssim 0.5$, and followed up in X rays with
$\textsl{XMM-Newton}$ (Aghanim et al. 2011b). Note that at $z\gtrsim 0.2$
further evidence will be difficult to obtain from X rays alone, given their
bias toward the high central brightness proper to LEs.

Further developments have been recently stimulated and provided by the
increasing resolutions (currently around tens of arcsecs) attained in
observing the SZ effect with ground-based instrumentations such as
\textsl{MUSTANG} (Korngut et al. 2011), \textsl{SZA} (Reese et al. 2012), and
\textsl{CARMA} (Plagge et al. 2012). In the present context, we stress the
relevance of the pressure profiles for CC and NCC (disturbed) clusters
obtained with \textsl{Bolocam} by Sayers et al. (2013) that, within the data
precision, are in tune with our discussion in \S~4.5.

Summing up, existing observations lend support to our bimodal pressure
profiles and to the basic evolutionary picture from low-$z$ LEs to high-$z$
HEs envisaged by the Grand Design. The next testing grounds will involve
observing the SZ profiles from more individual clusters of the HE and LE
types over an extended range of redshifts. This still constitutes a laborious
or challenging proposition with the current instrumentation even at
$z\lesssim 0.5$, but still coming of age with new-generation instruments such
as \textsl{SPT} \textsl{MUSTANG}, \textsl{ACT}, \textsl{CARMA},
\textsl{Bolocam} and the upcoming \textsl{ALMA} (see
\texttt{http://www.almaobservatory.org/}). Meanwhile, a baseline at very low
$z\approx 0.02$ has been just provided by the highly resolved \textsl{Planck}
data concerning the nearby Coma cluster, discussed in \S~5.2.

\section{Beyond Thermal and Dynamical Equilibria}

Deviations from thermal pressure support are expected both in the outskirts
of LE and at the  center of  HE clusters. Next we discuss these issues in
turn.

\subsection{Turbulent support in the LE outskirts}

Recall from \S~2 that at the boundary of evolving LE clusters one expects
weakening shocks with Mach numbers  $\mathcal{M}^2\lesssim 3$ and decreasing
thermalization efficiency.  Correspondingly, relatively more kinetic energy
seeps through as given in the shock rest frame by $v_2^2/v_1^2 = (1/4 +
3/4\mathcal{M}^2)^2$, ranging from $1/16\approx 9.1 \%$ for
$\mathcal{M}^2>>3$, through $1/4$ for $\mathcal{M}^2=3$, and up to $1$ for
$\mathcal{M}^2=1$, see Cavaliere et al. (2011b) and Fig.~3. These conditions
become more and more conducive to trigger turbulence in the ICP fluid.

The turbulent contribution to equilibrium is conveniently described in terms
of the ratio $\delta\equiv \delta p/p$ of the turbulent to the thermal
pressure. The boundary normalization $\delta_R$ is consistently set at the
shock jump just by $v_2^2/v_1^2$, while an inward decay
$\delta(r)=\delta_R\,\exp[-(R-r)^2/\tilde \ell^2]$ is expected on the basis
of standard arguments. In fact, an inward decline of $\delta(r)$ on a scale
$\tilde \ell \sim \lambda_{\rm pp}\, (c_s/v_2)^{3/4}\, (R/\lambda_{\rm
pp})^{1/4}\approx 1/2$ Mpc is provided by the classic cascade from large
`eddies' driven at the macroscopic coherence length $\sim R/2$, then
fragmenting down to small eddies where dissipation becomes effective (see
Kolmogorov 1941, Monin \& Yaglom 1965, Inogamov \& Sunyaev 2003); recently,
the theory has been extended to subsonic but super-Alfvenic turbulence in
conditions of subdominant magnetic relative to thermal pressure (cf.
Petrosian \& East 2008, Brunetti \& Lazarian 2011).

Pleasingly, it turns out that the total pressure $p+\delta p\equiv p\,
(1+\delta)$ can be straightforwardly included in the hydrostatic equilibrium
and solved by the SM; the result reads
\begin{equation}
{p(r)\over p_R}={1+\delta_R\over 1+\delta(r)}\,\left\{1+{2\, G\, m_p\over 5\,
p_R^{2/5}\, (1+\delta_R)}\,\int_r^R{\mathrm d}x~{M(<x)\over x^2\,
k^{3/5}(x)}\, \left[{1+\delta_R\over
1+\delta(r)}\right]^{3/5}\,\right\}^{5/2}~.
\end{equation}
This corresponds to $p$ and $k$ replaced everywhere in Eq.~(9) by
$p\,(1+\delta)$ and by $k\,(1+\delta)$; correspondingly, $T$ is replaced by
an effective $T\, (1+\delta)$. All that is to be expected since turbulent
eddies concur with the truly microscopic thermal degrees of freedom toward
dispersing and ultimately dissipating the residual kinetic energy $\propto
v_2^2$ seeped through the shock. If turbulence is stirred, the thermal
pressure required for overall support of the ICP against the DM gravitational
field is correspondingly decreased.

Note that the intensity parameter $y(r)$ of the thermal SZ effect defined in
Eq.~(11) is then \emph{lowered} relative to the strict thermal equilibrium
expression Eq.~(9) by the explicit factor $1/(1+\delta)$. Thus the SZ effect
can provide a \emph{direct} probing of a deficit in thermal pressure,
implying a considerable turbulent component in cluster outskirts for
retaining equilibrium (Cavaliere et al. 2011b).

\subsection{Missing SZ effect in the core of the Coma Cluster?}

A case study of the conditions prevailing near the center of an HE cluster is
provided by Abell 1656, the very rich and closeby Coma Cluster. The case
arises when the recent high-sensitivity and high-resolution view in X rays
with \textsl{XMM-Newton} (Churazov et al. 2012) is compared to the SZ view
provided by the \textsl{Planck} Collaboration, with the rich data rebinned to
an effective $\sim 1'$ resolution (Ade et al. 2013). Fig.~12 also illustrates
the profile of the SZ effect expected on using the thermal pressure profile
inferred from the X-ray fits to $S_X(r)$ and $T(r)$ with the SM (see Table~1
for the values of $k_c$ and $a$; also Fusco-Femiano et al. 2011). The result
is expressed in terms of the equivalent Rayleigh-Jeans decrement $\Delta
T\equiv -2\, y\, T_{\rm cmb}$ of the CMB temperature $T_{\rm cmb}\approx
2.73$ K, for easy comparison with the \textsl{Planck} measurements as
presented by Ade et al. (2013, their Fig.~4).

Fig.~12 highlights a \emph{deficit} in the values of $|\Delta T|$ as measured
by \textsl{Planck}, relative to those expected from the X-ray observations.
The discrepancy appears to be sharp at the center and well beyond the current
uncertainties budget as presented by Ade et al. (2013), waiting for
confirmations from the second half of \textsl{Planck} data. Such a SZ vs.
X-ray mismatch goes also beyond the uncertainties affecting the entropy
parameters obtained from our X-ray fits.

The mismatch may be marginally alleviated if one relied on the smoother,
less-resolved X-ray data from \textsl{ROSAT} instead of \textsl{XMM-Newton};
the latter is particularly sensitive to clumpiness effects, so biasing high
the brightness along with the apparent baryon fraction as discussed by
Simionescu et al. (2011) and Churazov et al. (2012). On the other hand, a
similar SZ vs. X-ray mismatch has been obtained with quite different fitting
tools by Ade et al. (2013). Thus the tension appears to be
\emph{model-independent}, and calls for a physical explanation.

The missing SZ is hard to account for in terms of an overall shape of Coma
ICP compressed along the l.o.s., given that it appears to feature, if
anything, some ($\lesssim 10\%$) elongation, see De Filippis et al. (2005).
On scales of $10^2$ kpc the presence in the ICP of substantial azimuthal
substructure adding to ongoing inner shocks (e.g., Ade et al. 2013;
Simionescu et al. 2013) may contribute to bias the local X-ray densities with
the attendant temperatures, hence the average X-ray pressure. A similar bias
might be induced on scales of some $10$ kpc by clumpiness (Simionescu et al.
2011) and fluctuations (e.g., Khedekar et al. 2013); however, given the
constraints set by Churazov et al. (2012) on the density fluctuations in the
central Coma, we expect such effects to be limited to some $5\%$.

Thus we are led to discuss whether the mismatch may be traced back to a truly
diffuse non-thermal pressure contribution $\delta p$ from turbulence or
suprathermal electrons, adding to the thermal $p$ toward overall hydrostatic
equilibrium as discussed in \S~5.1; then relative to the total pressure $\hat
p=p\, (1+\delta)$, the thermal component sampled by the SZ effect will be
lower on average. Note that the condition $n\, T\, (1+\delta)\approx$ const
may be read as $n\times T\, (1+\delta)\approx$ const implying an enhanced
kinetic temperature; this leads to broadening, shifting, and enhanced
excitation of X-ray spectral features, which have been proposed as marks of
non-thermal conditions (see Inogamov \& Sunyaev 2003, Sayers et al. 2013). In
this context, the SM formalism offers the extra gear of including such an
additional non-thermal component $\delta$ in the equilibrium condition
Eq.~(12) taken from Cavaliere et al. (2011a,b). For Coma a detailed shape
$\delta(r)$ is not required, and to a good approximation the main outcome is
to \emph{recalibrate} the thermal pressure to read $p\propto
(1+\delta)^{-1}$; resolving the tension between the SZ vs. the X-ray data
requires $\delta\approx 15-20\%$. The outcome is illustrated in Fig.~12 by
the solid line; we remark that while the SZ profile from the SM has not been
derived from a formal fit, yet it turns out to represent well the
\textsl{Planck} data over their whole radial range.

In a nutshell, the thermal electron pressure is related to the total
equilibrium pressure $\hat{p}$ by
\begin{equation}
p_e\approx {0.52\,\hat{p}\over 1+\delta}~.
\end{equation}
With $\delta\approx 15-20\%$, this boils down to $p_e\approx 0.45-0.42\,
\hat{p}$, definitely \emph{lower} than the level $0.52\, p$ pointed out in
\S~4.5. Note that sensible variations in the average ICP metallicity
$Z\approx 0.4\pm 0.03$ measured in Coma by Sato et al. (2011) would bias the
electron pressure inferred from the X-ray bremsstrahlung radiation by less
than a few percents, as discussed by Churazov et al. (2012). The strange case
of the missing SZ effect in Coma shows that also the inner volume of HE
clusters is likely to include an appreciable non-thermal component, though of
a possibly different nature from the outskirts' of LEs.

\subsection{The physical nature of non-thermal pressure in inner Coma}

Next we discuss the nature of such an inner \textit{non-thermal} pressure
contribution $\delta p$ to the overall equilibrium.

Ongoing turbulence, originated by recent mergers that drive turbulent wakes
and instabilities in the weakly magnetized ICP constitutes an attractive
contributor in view of its direct link to the primary merger energetics. Such
a turbulence has been widely discussed by many authors as a source of
velocity and density fluctuations (e.g., Nagai et al. 2007, Vazza et al.
2010, Iapichino et al. 2011); it is widely held to accelerate with moderate
efficiency supra-thermal electrons in the plasma to mildly relativistic
energies giving rise to steep energy distributions as discussed by
Schlickeiser et al. (1987), Sarazin \& Kempner (2000), Blasi et al. (2007),
and Brunetti \& Lazarian (2011). However, in Coma the density fluctuations
caused by ongoing subsonic turbulence have been constrained by Churazov et
al. (2012, see their \S~5.2 and 5.3) to be less than $5\%$ on scales $30-300$
kpc. The corresponding indirect estimates of current turbulent velocities
$\lesssim 450$ km s$^{-1}$ would fall short of providing the additional
pressure required to relieve the SZ vs. X-ray tension. The actual turbulence
velocities will be directly probed with the upcoming \textsl{ASTRO-H} mission
(\texttt{http://www.astro-h.isas.jaxa.jp/}).

Cosmic-ray protons are attractive diffuse contributors (e.g., Pfrommer et al.
2005), since their energy is longlived and can be stored within a cluster.
However, in Coma their overall energy density has been bounded to be less
than a few $10^{-2}$ of the thermal pressure by radio and $\gamma$-ray
observations (cf. Ackermann et al. 2010, Bonafede et al. 2011). On the other
hand, cosmic rays may still play a role as injectors of secondary electrons,
to be subsequently accelerated by turbulence and shocks in the ICP (as
discussed by Brunetti et al. 2012).

Thus we discuss the option offered by relativistic and trans-relativistic or
suprathermal electrons. Those with Lorentz factors $\gamma\gtrsim 10^3$ in
the diffuse magnetic field $B\approx$ a few $\mu$G measured in Coma emit the
large-scale synchrotron radiation observed at $\nu\gtrsim 30$ MHz in the form
of the classic Coma radiohalo, see Govoni et al. (2001) and Brunetti et al.
(2012). Based on the halo shape discussed in the last reference, the
pressures of the magnetic field and of the energetic electrons appear to be
effectively coupled to the dominant thermal population, as pointed out by
Brown \& Rudnick (2011) and Bonafede et al. (2011). The integrated radio
power of several $10^{40}$ erg s$^{-1}$ implies a relativistic energy density
of order $10^{-16}$ erg cm$^{-3}$ (cf. Giovannini et al. 1993, with
parameters updated). Although the corresponding pressure value is
substantially smaller than the required $\delta p\approx 0.15\, p\approx$
several $10^{-12}$ erg cm$^{-3}$, relativistic electrons can point to
interesting candidates provided their energy distribution extends steeply
toward a lower end $\gamma_1\lesssim 10^2$.

Such an extension is consistent with the radio spectrum retaining a slope
$\alpha\approx 1.2$ or somewhat steeper, as observed down to frequencies
$\nu\approx 31$ MHz (see Henning 1989); the corresponding electron
distribution is to rise toward low energies as $\gamma^{-s}$ with slope
$s\equiv 2\alpha+1\approx 3.4$. Existing data (reported by Henning 1989) also
show that at lower frequencies the radio flux in Coma is still sustained, and
may even feature a steeper component, as found in other clusters (e.g., van
Weeren et al. 2012); \textsl{LOFAR} will soon clear the issue (see
\texttt{http:/www.lofar.org/}).

The amount of \emph{non-thermal} pressure implied by the above electron
population may be estimated as $\delta p\approx \gamma_1\,m_e c^2\, n_{\rm
rel}(\gamma_1)/3\propto \gamma_1^{2-s}$, and refined with the full
expressions for mildly relativistic electrons given by En{\ss}lin \& Kaiser
(2000, their Appendix A). Based on the value $2\times 10^{40}$ erg s$^{-1}$
of the radiohalo luminosity at $100$ MHz and the profile given by Brunetti et
al. (2012), a non-thermal contribution $\delta p/p\approx 15\%$ would indeed
obtain if a straight electron distribution extended down to $\gamma_1\sim$ a
few.

On the other hand, a slope sustained against the fast Coulomb losses (e.g.,
Sarazin 1999, Petrosian \& East 2008) requires such electrons not to be drawn
from the thermal pool, but rather to have been injected over a few $10^7$ yr
by the action of mergers, or from AGNs (like the current sources associated
with NGC 4869 and NGC 4874), or by cosmic-ray interactions as discussed by
Brunetti et al. (2012). These electrons are widely held to be accelerated via
turbulence and low-$\mathcal{M}$ shocks, recently driven by mergers already
on the way of dissipating, so as to meet the constraints recalled above from
Churazov et al. (2012). We stress that similar merging events over timescales
of Gyrs are \emph{independently} required for providing the top level
$k_c\approx 500$ keV cm$^2$ of the central entropy measured in Coma.

On the other hand, a `silent pool' of cooling electrons with $\gamma \sim
10^2$ can be replenished and piled up since their lifetimes against Coulomb
and synchrotron losses top at about $1$ Gyr (cf. Sarazin 1999). With a
cumulative density $n\sim 10^{-7}$ cm$^{-3}$ resulting from several recent
mergers, these electrons can yield a non-thermal contribution $\delta\approx
15\%$. Their synchrotron and relativistic bremsstrahlung radiations would
easily escape detection (Sarazin 1999, Sarazin \& Kempner 2000), while their
collective contribution to pressure is \emph{probed} just through the thermal
SZ effect. Note that sustaining such a non-thermal pool requires from mergers
a considerable energy dissipation in acceleration, though far less than the
thermal dissipation.

If the energy distribution extends down to $\gamma_1 \sim$ a few, direct
evidence of trans-relativistic electrons may be provided; in fact, their
density scaling as $n\propto \gamma_1^{1-s}$ is up to $\sim 10^{-5}$
cm$^{-3}$, sufficient to gauge the low-$\gamma$ electron population via the
tail of the SZ effect at very high frequencies $\gtrsim 1$ THz, and the
accompanying displacement of the thermal null at $217$ GHz (see also Rephaeli
1995). Such features in the SZ spectrum are within the reach of sensitive
instrumentation like \textsl{ALMA} (see
\texttt{http://www.almaobservatory.org/}).

In conclusion, the intriguing physical conditions indicated by the inner ICP
in the Coma cluster apparently include both a \emph{thermal} and a
\emph{non-thermal} component, to be probed via three observational channels
across the electromagnetic spectrum: the bremsstrahlung emission in X rays,
the thermal and relativistic SZ effects in microwaves, and the diffuse
synchrotron radiation in the radio band.

\subsection{Outer breakdown of thermal and dynamical equilibria}

Adding to any non-thermal contribution to ICP macroscopic equilibrium as
described above, a specific microscopic breakdown may involve the electron
temperature relative to the ions' downstream the boundary shocks. As
anticipated in \S~1, this occurs because the protons share their momentum and
energy over a few mean free paths $\lambda_{\rm pp} \sim 10\, (k_B\, T/ 5
\,\mathrm{keV})^2$ $(n/10^{-3}\,\mathrm{cm}^{-3})^{-1}$ kpc, while the
initially cold electrons follow suite to local thermal equilibrium at a
common $T$ over a considerably longer scale $40\, \lambda_{\rm pp}$ (see the
pioneering assessment by Zel'dovich \& Raizer 1967, and the estimates focused
to the cluster context by Wong \& Sarazin 2009). Such microscopic
disequilibria can also cause SZ deficits, but at a few percent levels; these
are lower than the effects of non-thermal overpressures $\delta p$ affecting
the outskirts of LE clusters discussed in \S~5.1, while becoming more
relevant behind the strong boundary shocks that prevail in HE clusters.

On the macroscopic side of the ICP equilibrium, a strongly ellipsoidal
overall geometry looked at along the minor axes tends to yield lower values
of $y$ than expected from lines-of-sight extending out to the major axis
length (as pointed out by Korngut et al. 2011, Reese et al. 2012). Here the
linearity of SZ effect plays against, whereas the X-ray brightness is anyway
dominated by the central, likely more spherical density peak. The resulting
discrepancies are estimated at $10\%$ levels by the above authors.

While spherical hydrostatic equilibrium (possibly supplemented by non-thermal
components) provides the simplest benchmark to describe average features of
the ICP state, it is likely to progressively break down toward the virial
radius and beyond. Over limited angular sectors, this is what we have
discussed in \S~4.3 (see the observations referenced therein) as to the
inflows of DM and outer intergalactic gas channeled into the developing
cluster along large-scale filaments. These not only break to some degree the
azimuthal symmetry, but also carry down outer gas conceivably preheated by
lateral compressions and external shocks, yielding at the virial boundary
effective Mach numbers lower than from direct spherical infall (see Vazza et
al. 2009, Kravtsov \& Borgani 2012).

On a Mpc scale, extreme mergers with mass ratios close to 1:1 may trigger
conditions of severe ICP disequilibrium such as observed in A754 (Macario et
al. 2011) and A2146 (Russell et al. 2010), or outright disruption like in
MACS J0025.4-1222 (Brada\v{c} et al. 2008) and in the prototypical 1E0657-56
(the `Bullet Cluster', Clowe et al. 2006). This is the domain where
high-resolution simulations of templates for individual cluster collisions
are of key relevance (as discussed by Kravtsov \& Borgani 2012).

\section{Summary and Outlook}

In the above \emph{Report} our thrust has been toward providing a coherent
account of the complex physics of the intracluster plasma, as it emerges when
hundreds or thousands of galaxies cluster within a few Mpcs. These conditions
cause the ICP to \emph{react} in different ways to the DM gravity pull by
means of different rates of entropy production and of radiative losses. We
have been motivated by the clear signals that passive ICP self-similarity $-$
the simplest unifying scheme $-$ is broken by the observables and their
profiles inside and outside $r\sim R_{500}\simeq R/2$; specifically, it
cannot cover the cool-core together with the non-cool-core clusters, and even
less the groups together with the cluster lot. In the quest for a more
specific and effective pattern, we have focused onto the conversion of
gravitational into thermal energy with the associated entropy production, and
have pursued the following steps. First, relate the entropy produced in the
ICP to the development stages of the dark matter halos; second, discuss on
worked issues the leading role that the entropy so produced plays in
modeling/fitting the plasma observables; third, provide a consistent view of
the long-term cluster evolution, and highlight the relationships to group
conditions.

$\bullet$ As to our first step, we have described how the basic entropy
\emph{pattern} $k(r)=k_c+k_R\, (r/R)^a$ arises from the two stages recently
recognized in the DM halo formation process. A central floor $k_c\sim 10^2$
keV cm$^2$ is set by the early violent collapse that includes the plunging of
several major mergers. This is started by an initial cosmogonic perturbation
gone non-linear, and ends up with condensing intergalactic baryons to core
densities $n\sim 10^{-3}$ cm$^{-3}$ while heating them to a few keVs. The
subsequent, prolonged development stage includes accelerated radiative
cooling that at the centers works to erase the initial value of $k_c$. But
major energy injections from residual deep mergers and from violent AGN
outbursts will intermittently raise it again to values around $10^2$ keV
cm$^2$, enough to slow down or even reset the cooling.

Meanwhile, the entropy outer ramp $k\propto r^a$ with slope $a\sim 1$ grows
from the inside-out for several crossing times $R/\sigma\sim 1$ Gyr, as long
as cold outer gas supersonically infalls at a brisk rate into the developing
halo. The baryons are strongly shocked and halted in a layer around the
current virial radius, where entropy is continuously produced over several
Gyrs, and stratified into the ramp extending up to the boundary value
$k_R\sim$ a few $10^3$ keV cm$^2$. Eventually, however, the inflows dwindle
away as they draw on the tapering wings of the initial perturbation over the
background density provided by the accelerating universe, and/or become
transonic as they are channeled down and preheated within filaments of the
large-scale cosmic web. Then the virial shocks weaken and the entropy
production slows down, with the outer ramp flattening out if not bending
over. Thus the radial entropy distribution carries a stratified record of the
second stage of the cluster formation history spanning several Gyrs.

$\bullet$ As to our second step, the above processes set the average radial
entropy run that in turn \emph{modulates} the ICP profiles of density $n(r)$,
temperature $T(r)$ and pressure $p(r)$ within the DM potential wells. Over
much of the cluster lifetime and spatial extent, this interplay is well
described by spherical hydrostatic equilibrium under thermal pressure, which
constitutes the primary, robust variable in our Supermodel (SM). Pressure not
only is prompt to recover in a few sound crossing times its monotonic course
after the impacts of minor to intermediate mergers, but also is directly
\emph{keyed} to the entropy run after our SM Eq.~(9). Finally, pressure is
open to direct measurements via the Sunyaev-Zel'dovich effect. The SM (with
its finite total mass and no central singularities in any of the ICP
observables) provides a handy fitting tool for the l.o.s.-integrated, $2$-D
profiles of pressure, as well as for the associated X-ray brightness and
temperature, all keyed to the \emph{same} $3$-D entropy spine. The fitting
procedure involves a few non-degenerate parameters, and dispenses with
delicate deprojections.

Such analyses show that many rich clusters with masses in the range
$10^{14-15}\, M_\odot$ can be divided into two main classes: LEs and HEs,
featuring low or high entropy, respectively. This description refers to both
the core levels $k_c$ (corresponding to cool-core and non-cool-core states)
and to the boundary values $k_R$. As a consequence, HEs feature flat
temperature and brightness profiles at the center, but still underdeveloped
outskirts especially at high $z>0.5$, with related low halo concentrations
$c<5$. LEs, instead, feature a middle peak in the temperature profile, with a
cooler center and an associated brightness spike; toward the outskirts their
temperature decline steepens down, particularly at low $z$. Relatedly, they
also feature high concentrations $c\gtrsim 6$, indicative of long lifetimes.
In terms of the pressure profiles $p(r)$, we expect \emph{two} basic
templates applying to HEs and to LEs; while each is reasonably definite
within its class, the former is centrally quite shallower as indeed has been
recently observed (see discussions in \S~4.4 and 4.5; also Lapi et al. 2012).
The need for two templates to describe the measured pressure profiles by
itself highlights that ICP self-similarity is systematically \emph{broken} in
the cluster body, beyond object-to-object variance. Our SM analyses in \S~4.3
offer a handle to understand how the observables change at $r\sim R_{500}$
and in different azimuthal sectors. This occurs particularly in low-$z$ LE
clusters where the entropy production is drying out, and correspondingly the
outer entropy slope $a$ tends to vanish (cf. \S~3, just after Eq.~10).

Groups with their masses $10^{13-14}\, M_\odot$ fit into this framework just
by way of their shallower DM potential wells. At centers the groups are
sensitive to moderate energy injections with associated entropy raise
produced during the first halo collapse, or contributed by AGN outbursts; in
the outskirts their shallower wells draw slower infall of intergalactic
matter and imply less entropy production. The resulting ICP distributions are
more like HEs in the center and more like LEs in the outskirts; so groups
break self-similarity \emph{\emph{twice}}, as it were.

$\bullet$ As to our third step, the above pieces of evidence mainly related
to low-$z$, X-ray observables suggest the following evolutionary picture,
that we dubbed cluster Grand Design. This envisages a basic course from young
unrelaxed HEs to old relaxed LEs, with the latter's entropy reduced on
comparable timescales both at the center due to radiative cooling, and in the
outskirts due to progressively weakening accretion shocks. Such a basic
course may be halted and temporarily reversed by the impacts of large deep
mergers, that can remold an LE into an HE state. These events may cause
complex substructures in the ICP including internal shocks, cold fronts,
central sloshing, and even outright disruptions of equilibrium (for an
excellent review see Markevitch \& Vikhlinin 2007).

This evolutionary picture has been tested out to higher and higher $z$ by
looking first at stacked Sunyaev-Zel'dovich data of intermediate redshift
clusters; these show the HE population to increasingly dominate for $z\gtrsim
0.2$, as we expected. Supporting evidence is provided by the X-ray
observations of \textsl{Planck}-discovered clusters at $z\sim 0.5$, which
show pressure and density profiles consistent with the HE template. At $z$
higher yet, a statistical testbed for the above evolutionary picture is being
provided by the power spectrum of the unresolved SZ effect integrated over
redshift and over the evolving cluster mass distribution including groups
(e.g., Shaw et al. 2010, Efstathiou \& Migliaccio 2011). In Lapi et al.
(2012) we have shown how the tight constraints at multipoles $\ell\lesssim
3000$ set by current observations of the integrated thermal SZ effect with
the \textsl{SPT} (Reichardt et al. 2011) are converging to indicate at the
relevant redshifts $z> 0.5$ a take over of HEs with their lower central
pressures. Moreover, the data body suggests a widespread presence of
non-thermal contributions to the inner ICP equilibrium in these high-$z$ HEs,
similarly to the conditions prevailing in the nearby Coma cluster. Additional
constraints to the integrated SZ effect from clusters will require high
sensitivities at resolutions $\ell>3000$, with good control of systematics.
To the purpose, it is also required a precise independent determination of
the normalization $\sigma_8\approx 0.8$ for the primordial cosmogonic power
spectrum at cluster scales (cf. Komatsu et al. 2011, Hinshaw et al. 2013), as
may be expected from the final release of the entire \textsl{Planck} dataset
(see \textsl{Planck} Collaboration 2013).

Other cosmological parameters may be sensitively probed through the evolution
of the cluster mass function $N(M,z)$, as pressed on particularly by
Vikhlinin et al. (2006, 2009). To derive the current gravitating masses $M$,
one may take advantage once again of hydrostatic equilibrium and X-ray
observables on inverting Eq.~(6) to yield
\begin{equation}
M_X(<r)=-{r^2\over G}\, {1\over m_p\, n}\, {{\rm d}p\over {\rm d}r}~,
\end{equation}
as proposed by Fabricant \& Gorenstein (1983), see also Sarazin (1988) and
references therein. However, a widespread concern revolves around systematics
affecting such determinations at levels exceeding $10\%$. The main such
systematics is constituted by non-thermal contributions to the outer ICP
equilibrium, as pointed out in several recent works and in particular by
Reese et al. (2012, see their Table 10). In this context, here we carry a
step further our discussion of \S~5.1 from recalling that non-thermal
contributions are particularly relevant to the outskirts of evolved LE
clusters.

To proceed, once again we take advantage of the asymptotic scaling laws for
the outer pressure profile given in \S~3 (see also Fig.~3). So it is seen
that $M_X(<r)\sim r\, T\, {\rm d}\log p/{\rm d}\log r\propto (5-2 a)\,
r^{7\,a/5-1}$ holds, which highlights how unphysical non-monotonicity may
easily arise with \emph{flat} entropy slopes $a\lesssim 0.7$. This in fact is
the case with a number of clusters observed by \textsl{Suzaku} out to the
virial radius (see Sato et al. 2012 and references therein). Our point is
that in such clusters weaker boundary shocks prevail and let relatively more
bulk inflow energy to seep through (as illustrated by Fig.~3), ready to drive
more turbulence. In such cases Eq.~(12) shows how the associated non-thermal
pressure can be straightforwardly included in the SM; not only this helps to
recover a realistic equilibrium with monotonically increasing total mass (and
baryonic fraction at the cosmic value, see Fusco-Femiano \& Lapi 2013), but
also leads to gauge the bias of the mass $M_X$ as reconstructed from X-ray
observations when the non-thermal contribution is ignored.

In such circumstances, from Fig.~13 it is seen how $M_X(<r)$, in the presence
of an ignored, outer non-thermal component of order $\delta_R\sim 10-20\%$
will deviate downward from the true mass by comparable amounts. For
turbulence decay scales $\tilde \ell<0.3\, R$, the reconstructed mass
$M_X(<r)$ will still retain an apparent non-monotonic behavior, not unlike
the results from several observations, including Kawaharada et al. (2010) in
three sectors of the much studied cluster A1689 (cf. their Fig.~8). On the
one hand, these studies provide useful lower bounds to the thickness of the
turbulent layer. On the other hand, accounting for the turbulence bias
constitutes a key point to resolve the tension between weak lensing and X-ray
masses (e.g., Nagai et al. 2007, Lau et al. 2009, Meneghetti et al. 2010),
and to derive precise cosmological parameters from statistics of cluster
masses via the fast X-ray observations (see Vikhlinin et al. 2009).

In this context, we stress that SZ observations not only can crosscheck the
bias of $M_X$ (as originally proposed by Cavaliere et al. 2005, Ameglio et
al. 2007), but also may provide a \emph{measure} thereof in terms of the
X-ray vs. SZ pressure indicator $b\equiv p_X/p_{SZ}-1$, as defined by
Khedekar et al. (2013). In fact, after Eq.~(14) which applies with a partial
non-thermal support, the mass bias simply works out to
\begin{equation}
b = \delta~,
\end{equation}
which may be used to check and correct the X-ray mass estimates.

Control over such a mass bias will be mandatory to take full advantage of
upcoming, large X-ray surveys like $\textsl{e-ROSITA}$ (see
\texttt{http://www.mpe.mpg.de/ eROSITA}) in measuring the parameter
$w_\Lambda$ that modulates the dark energy equation of state, and in gauging
its evolution with redshift. On a longer perspective, added value will be
provided by independent measurements of weak lensing from wide-area cluster
surveys as expected from the \textsl{Euclid} mission (see
\texttt{http://www.euclid-ec.org/}).

As anticipated in \S~2, this \textit{Report} reviews a novel approach and a
number of worked issues to understand why and how self-similarity or simple
regularity break down in the ICP at around $R_{500}$, differently from the
DM. In closing, we raise a unifying issue: up to what point do the
collisional ICP distributions in rich clusters (discussed throughout the
present \emph{Report}) mimic the underlying collisionless DM profiles
(recalled in \S~2.1)? Paradoxically, the latter most closely resemble the ICP
distributions in LE clusters where collisional radiative interactions erase
the core left over by the early collapse. The resemblance goes to the point
that LEs are closely described in terms of a temperature profile $T(r)$
mirroring the DM velocity dispersion $\sigma^2(r)$ as given by $T(r)\propto
\sigma^2(r)$, see discussion in \S~3.2.

This challenging situation is best phrased in terms of a similarity between
the distribution of the ICP thermodynamic entropy $k=k_B T/n^{2/3}$, and the
featureless DM gravitational `pseudo-entropy' $K\equiv \sigma^2/\rho^{2/3}$
widely used in the relevant literature (e.g., Taylor \& Navarro 2001,
Faltenbacher et al. 2007, Lapi \& Cavaliere 2009 and references therein). In
fact, the ICP entropy run $k(r)\propto r$ applies throughout the LE cluster
bodies, and turns out to closely resemble the DM pseudo-entropy run
$K(r)\propto r^{1.25}$. This holds despite $k(r)$ being produced by
short-range collisional processes that lead to heating and cooling, while
$K(r)$ originates from long-range, gravitational interactions leading to wide
orbital mixing (from Lynden-Bell 1967 through Bertschinger 1985 to Lapi \&
Cavaliere 2011).

In point of principle, both entropies express the `occupation' of the
appropriate phase-spaces; in point of fact, they provide the operational link
between the density and the velocity dispersions $\sigma^2$ or $T$. As shown
by Cavaliere et al. (2009), the link enters in closely similar ways the
respective equilibria governed by the hydrostatic equation for the ICP, and
by the analogous Jeans equation for the cold DM. It will be worth considering
how the picture would change in the increasingly entertained warm DM
scenario.

\bigskip
\bigskip

Work supported by MIUR through the PRIN 2011/2012. We thank the anonymous
referee for constructive and helpful comments. We are grateful to R.
Fusco-Femiano for fruitful collaborations and stimulating discussions
concerning the material of this \emph{Report}. We acknowledge helpful
exchanges with G. Brunetti, E. Churazov, L. Danese, G. De Zotti, M.
Kawaharada, M. Massardi, P. Mazzotta, M. Migliaccio, S. Molendi, D. Nagai, P.
Natoli, Y. Rephaeli, and P. Salucci. A.L. thanks SISSA for warm hospitality.

\bibliographystyle{elsarticle-num}

\clearpage
\begin{figure}
\center
\includegraphics[height=20cm]{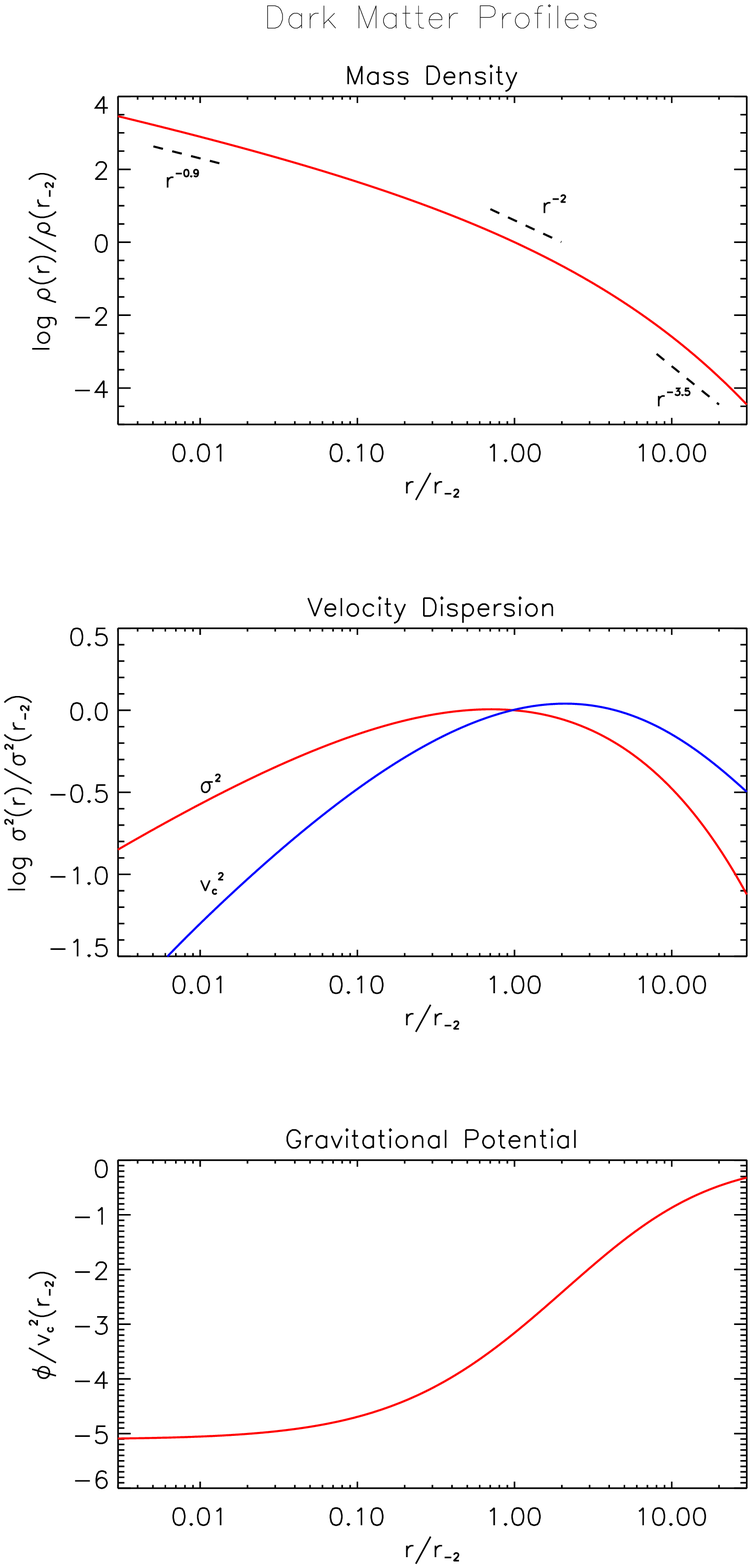}
\caption{Spherically-averaged profiles of cold DM halos. Top panel: density
profile, with typical slopes (see Eq.~4) in the inner, middle, and outer
regions illustrated by the dashed lines. Middle panel: profiles of $1$-D
velocity dispersion (red) and circular velocity (blue). Bottom panel: profile
of the related gravitational potential. In the first two panels the profiles
are normalized at the radius $r_{-2}$ where the logarithmic density slope
equals $-2$, whereas in the third panel the profile is normalized to the
virial velocity squared.}
\end{figure}

\clearpage
\begin{figure}
\center
\includegraphics[height=15cm]{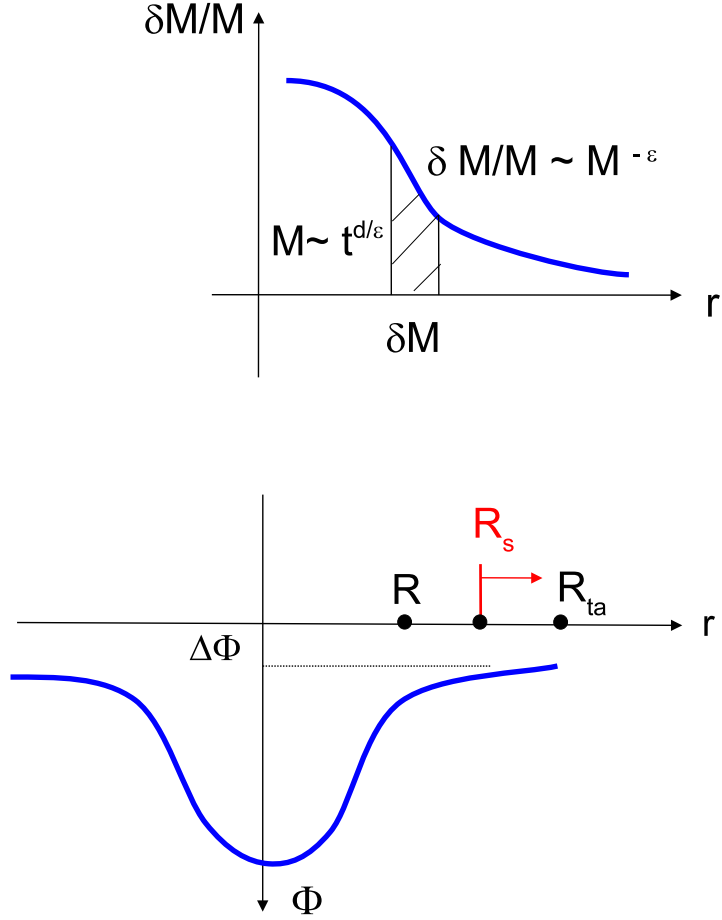}
\caption{Schematics of DM halo formation from a scale-invariant
density perturbation with shape $\delta M/M\propto M^{-\epsilon}$ (top panel),
with the corresponding gravitational potential (bottom panel) that governs
intergalactic gas infall, see \S~2.3. As the outskirts develop,
mass shells $\delta M$ drawn from the wings of the perturbation
accrete onto the enclosed mass $M$; the infall starts at the turnaround radius
$R_{\rm ta}\approx 2\, R$ toward the shock radius $R_s\approx R$. The growth
of the mass $M\propto t^{d/\epsilon}$ is modulated by the
perturbation shape via $\epsilon\gtrsim 1$, and by the background cosmology
via the linear growth factor $D\propto t^d$ with $d\approx 2/3-1/2$. During
the late growth, the outer potential drop $\Delta \Phi/v_R^2\approx
[1-(R_s/R_{\rm ta})^{3\epsilon-2}]/(3\epsilon-2)$ lowers as $\epsilon$
increases in the perturbation wings. In addition, as the accretion rates
$\dot M/M\approx d/\epsilon\, t$ subside, the ram pressure of the infalling
gas decreases and the shock position $R_s$ slowly outgrows $R$, further
lowering the ratio $R_s/R_{\rm ta}$ and the effective $\Delta\Phi$. As
a result, the accretion shocks weaken, with less entropy
produced and relatively more kinetic energy seeping through the shock
(see also Fig.~3).}
\end{figure}

\begin{figure}
\center
\includegraphics[height=10cm]{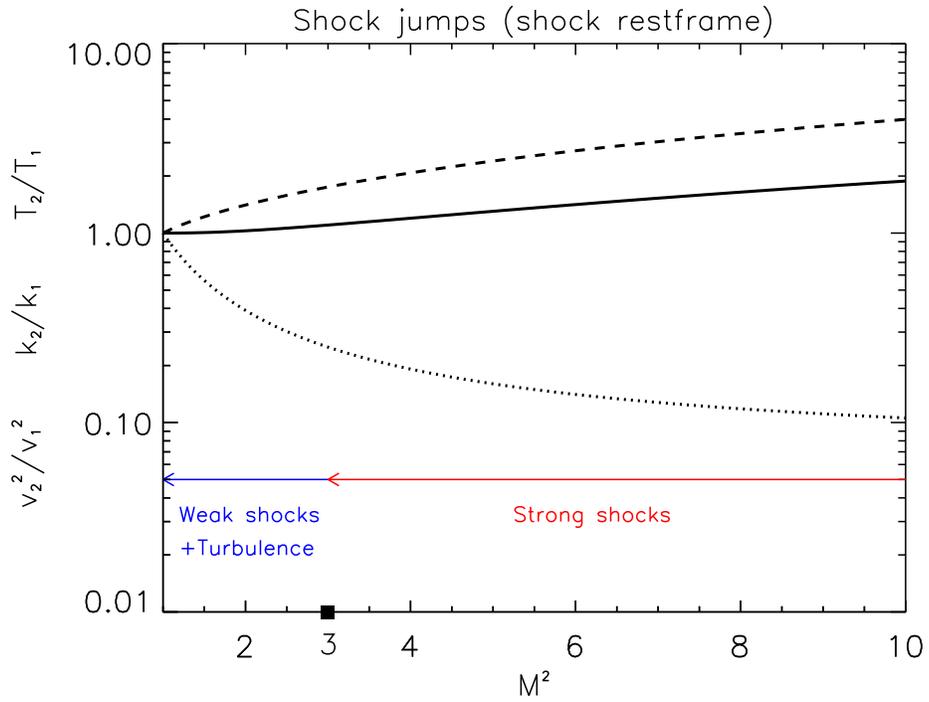}
\caption{Plots of the jumps across the shock for the entropy $k_2/k_1$ (solid
line), the temperature $T_2/T_1$ (dashed line), and the residual kinetic
energy $v^2_2/v^2_1$ (dotted line) as a function of the squared Mach number
$\mathcal{M}^2$; the divide between `\emph{strong}' and `\emph{weak}' shocks
(as effective or ineffective entropy producers) is around
$\mathcal{M}^2\approx 3$; note that the bound $v_2^2/v_1^2\gtrsim 6.3\%$
always applies, see \S~2.3. During a cluster's evolution, the outskirts
condition progresses from right to left, i.e., from strong to weak shocks.
The corresponding detailed expressions are discussed in Lapi et al. (2005)
and given explicitly by Cavaliere et al. (2011a).}
\end{figure}

\clearpage
\begin{figure}
\center
\includegraphics[height=10cm]{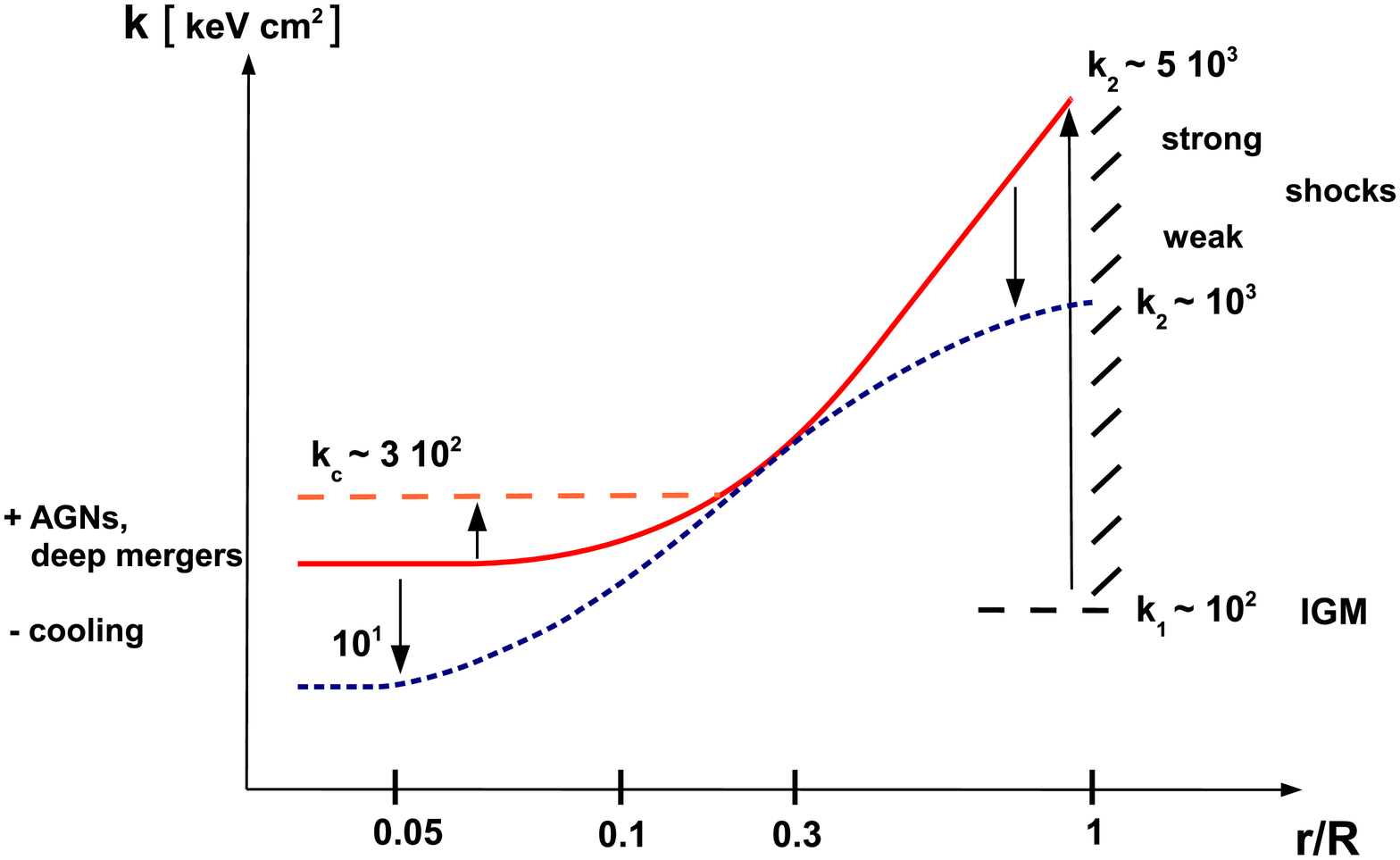}
\caption{The schematics illustrates our fiducial patterns for the ICP entropy
distribution $k(r)$. In the basic pattern (central floor plus ramp; red solid
line), entropy is raised at the boundary from intergalactic values $k_1\sim
10^2$ keV cm$^2$ to high outer levels $k_2\sim$ several $\times 10^3$ keV
cm$^2$ by strong boundary shocks. As the outskirts develop, the shocks weaken
and the outer level lowers to $k_2\lesssim 10^3$ keV cm$^2$; meanwhile, the
central entropy is eroded by radiative cooling down to low time-integrated
levels $k_c\approx 10^1$ keV cm$^2$ (blue dotted line). On the other hand,
blastwaves driven by deep mergers may reset the central levels $k_c$ up to
several $\times 10^2$ keV cm$^2$, and easily spread it out in the form of an
extended hot spot (orange dashed line).}
\end{figure}

\clearpage
\begin{figure}
\center
\includegraphics[height=15cm]{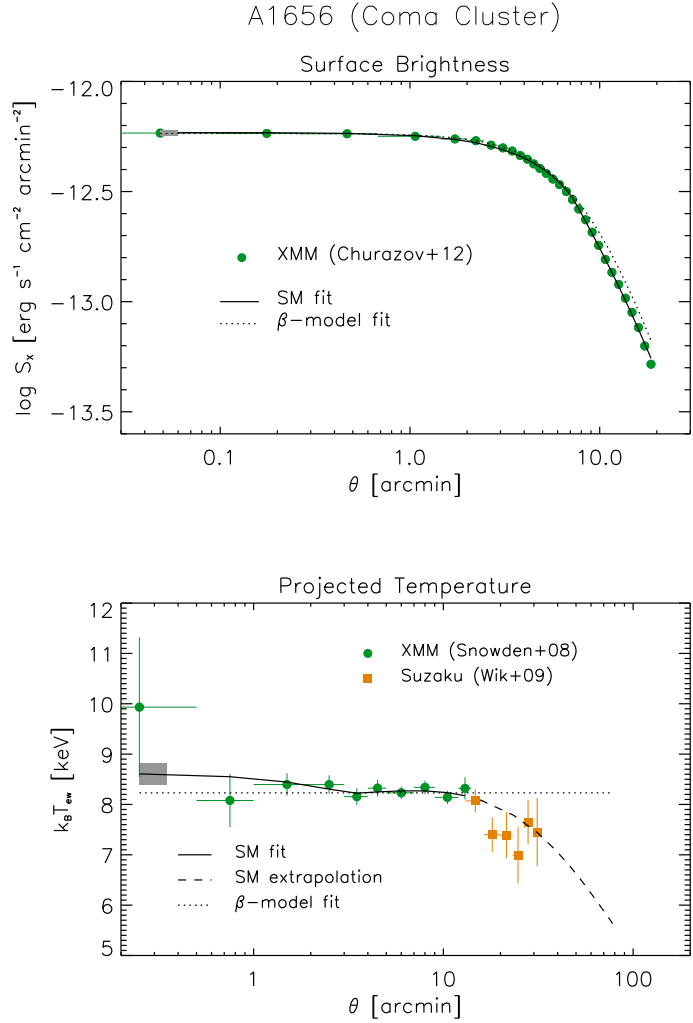}
\caption{Profiles of X-ray observables in the NCC cluster Abell 1656 (Coma
Cluster). Top panel: X-ray surface brightness; green circles refer to the
\textsl{XMM-Newton} data by Churazov et al. (2012), dotted line shows the
$\beta-$model fit from these authors, while solid line illustrates the
Supermodel outcome. Bottom panel: Projected emission-weighted temperature;
green circles refer to the \textsl{XMM-Newton} data by Snowden et al. (2008),
orange squares to the \textsl{Suzaku} data by Wik et al. (2009); solid line
is the SM outcome, with the dashed line representing its extrapolation out to
the virial radius $R\approx 2.2$ Mpc; dotted line illustrates the fit with an
isothermal $\beta$-model. In both panels the shaded areas show the associated
$2$-$\sigma$ uncertainties of the fits.}
\end{figure}

\clearpage
\begin{figure}
\center
\includegraphics[height=15cm]{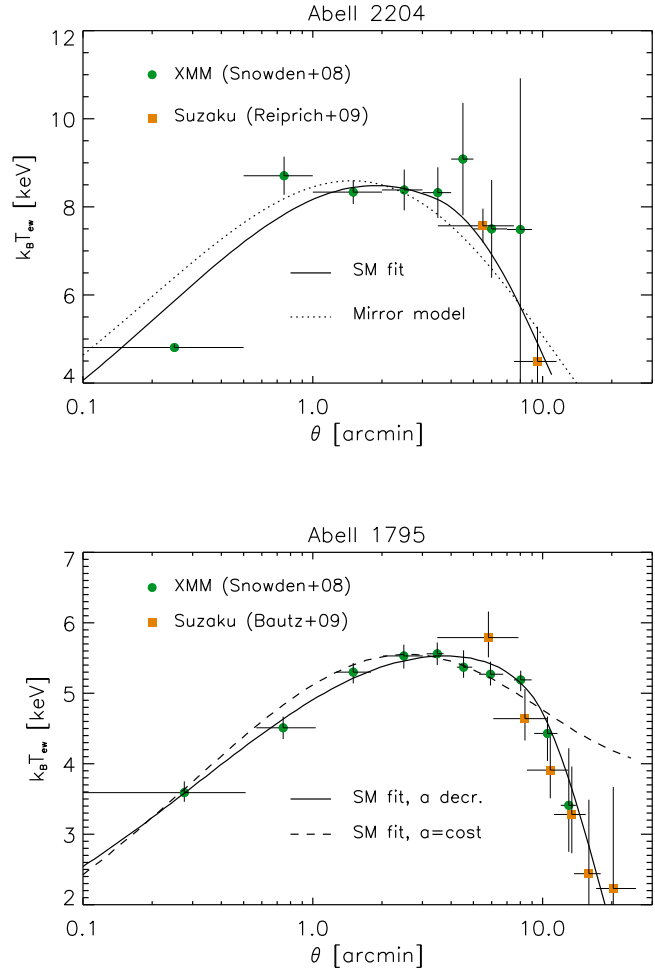}
\caption{Projected profile of emission-weighted temperature in two CC
clusters. Top panel: Abell 2204. Green circles refer to the
\textsl{XMM-Newton} data by Snowden et al. (2008), orange squares to the
\textsl{Suzaku} data by Reiprich et al. (2009). Solid line illustrates the
Supermodel outcome, while dotted line represents the `mirror' model discussed
in \S~3.2. Bottom panel: Abell 1795. Green circles refer to the
\textsl{XMM-Newton} data by Snowden et al. (2008), orange squares to the
\textsl{Suzaku} data by Bautz et al. (2009). Solid line illustrates the
Supermodel outcome with an entropy profile flattened down in the outskirts as
discussed in \S~4.3 and observed by Walker et al. (2012), while dashed line
represents the result computed with a constant entropy slope. Note the
finite central temperature levels, and the middle temperature maxima in
stark contrast with Coma (see Fig.~5, bottom panel).}
\end{figure}

\clearpage
\begin{figure}
\center
\includegraphics[height=12cm]{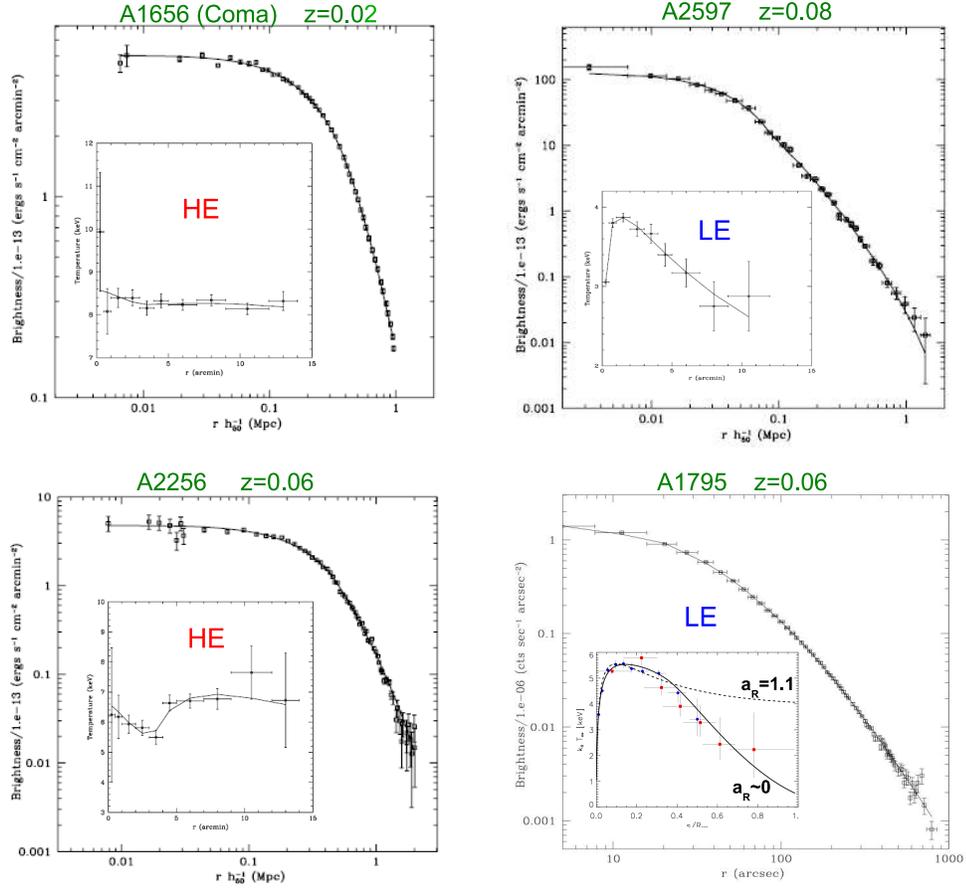}
\caption{The bimodality of HE vs. LE cluster is illustrated with a collection
of brightness and temperature profiles (main panels and insets, respectively)
for the HE clusters A1656 (top left panel) and A2256 (bottom left), together
with the LE clusters A2597 (top right) and A1795 (bottom right); lines show
the Supermodel fits. Note how the different values of the DM halo
concentration $c<5$ in HEs and $c\gtrsim 6$ in LEs reflect into the
position of the knee in the brightness profiles.}
\end{figure}

\clearpage
\begin{figure}
\center
\includegraphics[height=15cm]{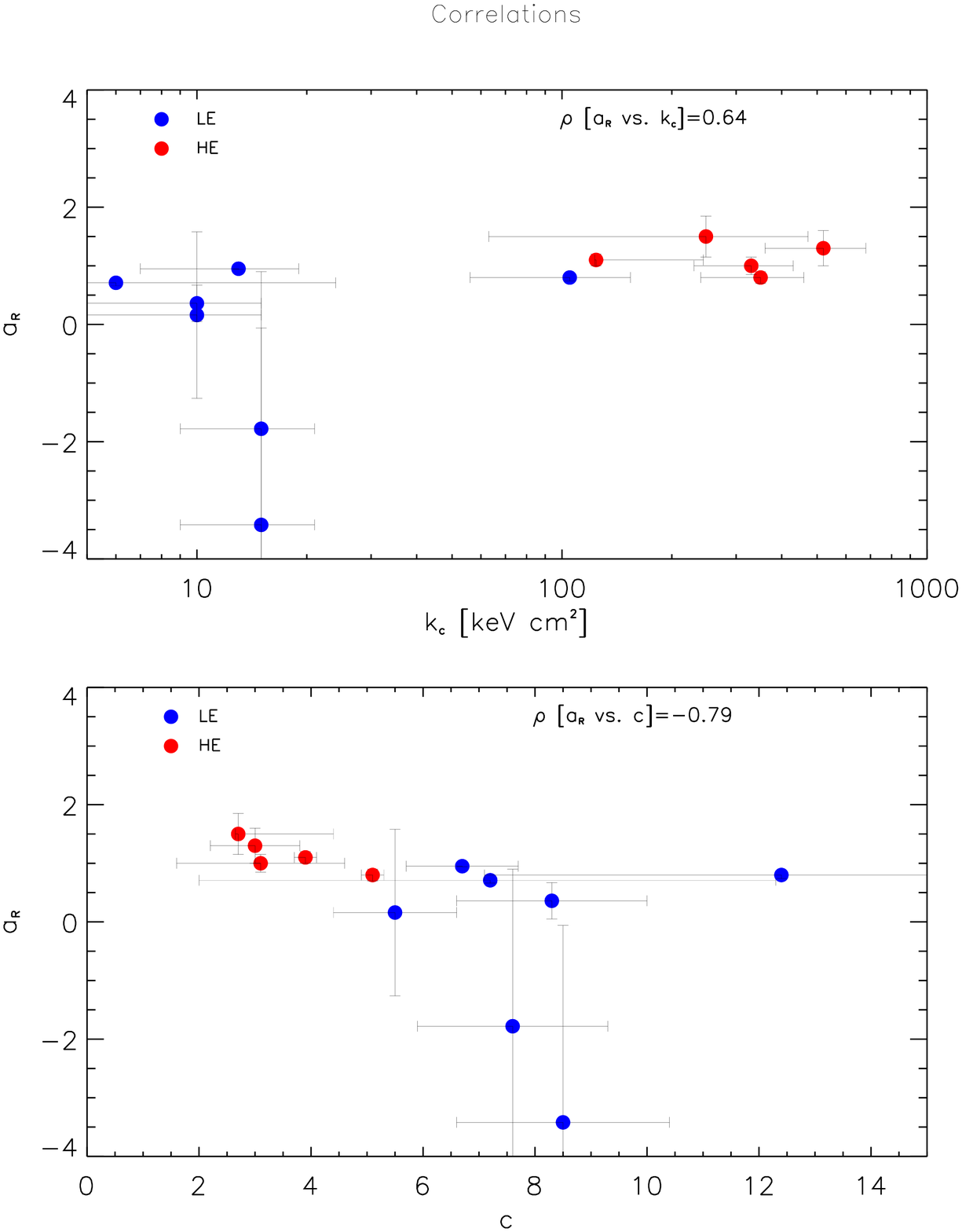}
\caption{Correlations between the ICP entropy and DM parameters for the
$12$ clusters listed in Table 1. Top panel: central entropy level $k_c$ vs.
outer entropy slope $a_R$. Dots illustrate the results from the Supermodel
analyses (red dots refer to HEs and blue dots to LEs). Bottom panel: the DM
concentration $c$ vs. the outer ICP entropy slope $a_R$; symbols are as
above. In both panels the Spearman's rank correlation coefficients $\rho$ for
the average data values are reported.}
\end{figure}

\clearpage
\begin{figure}
\center
\includegraphics[height=10cm]{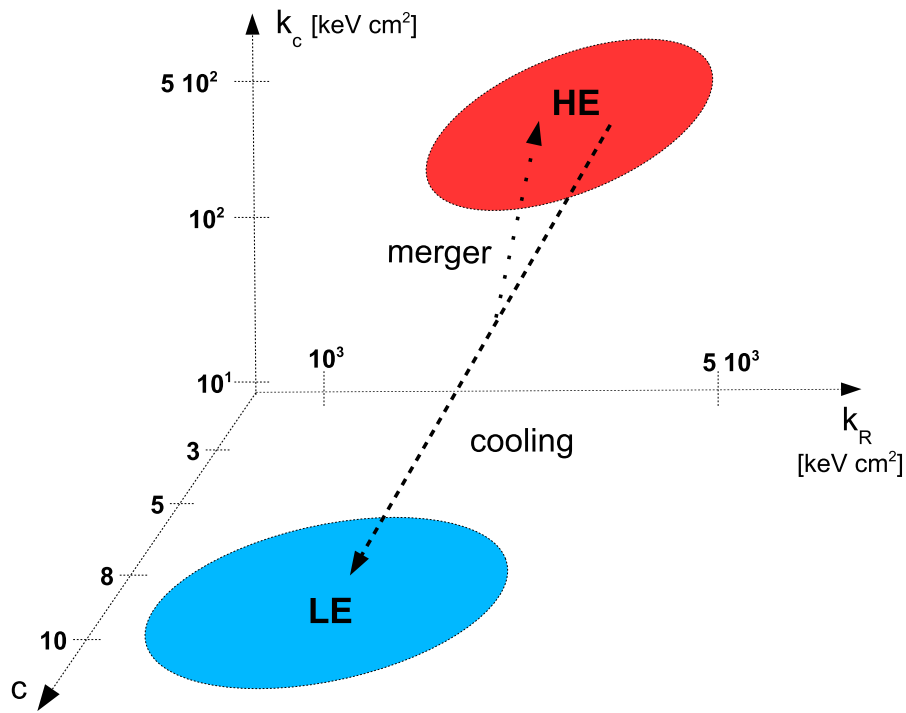}
\caption{A schematics illustrating the relationships among the cluster
classes after the Grand Design (see \S~4.2). The main trend proceeds from HE
to LE due to entropy erosion at the center by radiative cooling, and to
reduced entropy production at the boundary by weakening shocks. A deep merger
may occasionally halt and revert this course, originating a remolded HE
cluster with higher inner entropy.}
\end{figure}

\clearpage \clearpage
\begin{figure}
\center
\includegraphics[height=10cm]{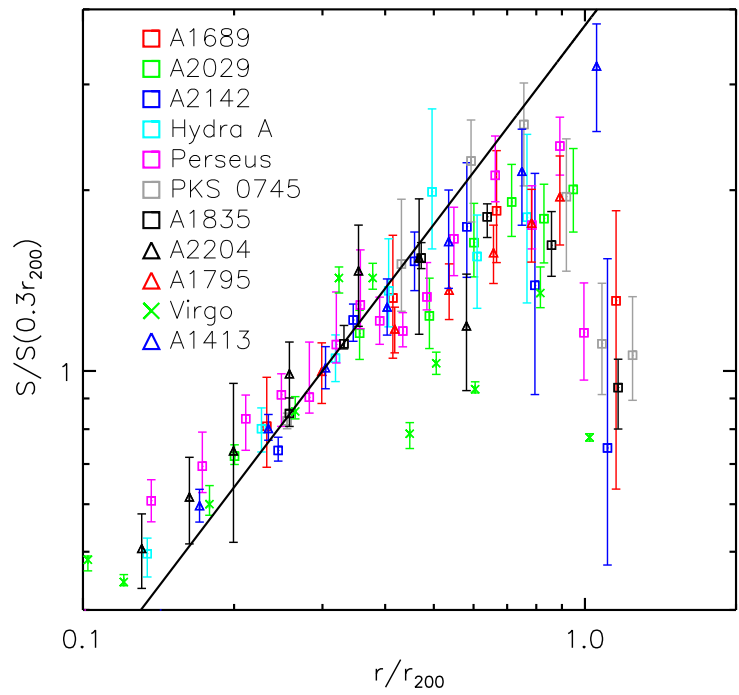}
\caption{Entropy profiles (normalized at $r=0.3\, R$) for a sample of
$11$ clusters observed in X rays. The solid black line shows the powerlaw
behavior $r^{1.1}$. [Credit: Walker et al. 2012].}
\end{figure}

\clearpage
\begin{figure}
\center
\includegraphics[height=10cm]{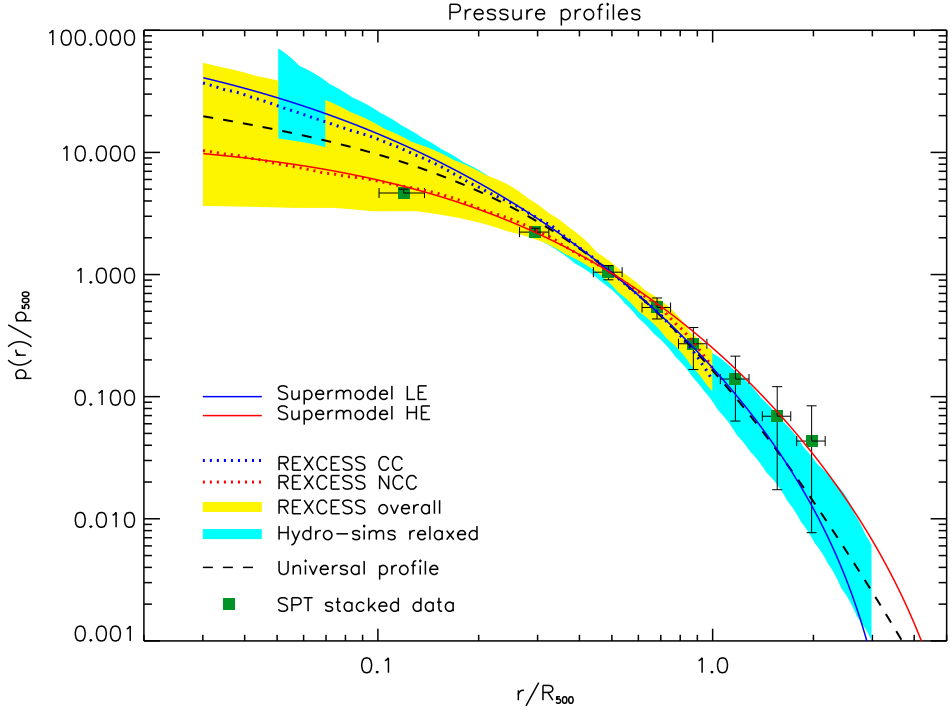}\\
\caption{Detailed profiles of ICP pressure normalized to $p_{500}$. Our SM
templates for HE and LE clusters are illustrated by the red and blue
solid lines, respectively (see \S~4.4 for parameter values). The yellow
shaded area illustrates the region covered by the low redshift ($z\lesssim
0.2$) clusters of the \textsl{REXCESS} X-ray sample; the dotted blue and red
lines refer to the average profiles separately for the subsamples of CC
(relaxed) and NCC (often disturbed) clusters, as defined by Arnaud et al.
(2010). The cyan shaded area illustrates the coverage by hydrodynamical
simulations of relaxed clusters. The dashed line represents the joint fit to
the observational and virtual data with the empirical `universal' profile
given in \S~4.4. The green squares represent stacked SZ observations of
higher redshift ($0.2\lesssim z\lesssim 0.4$) clusters with the \textsl{SPT}.
It is apparent the evolution toward an HE-like template as envisaged by the
Grand Design in \S~4.4.}
\end{figure}

\clearpage
\begin{figure}
\center
\includegraphics[height=10cm]{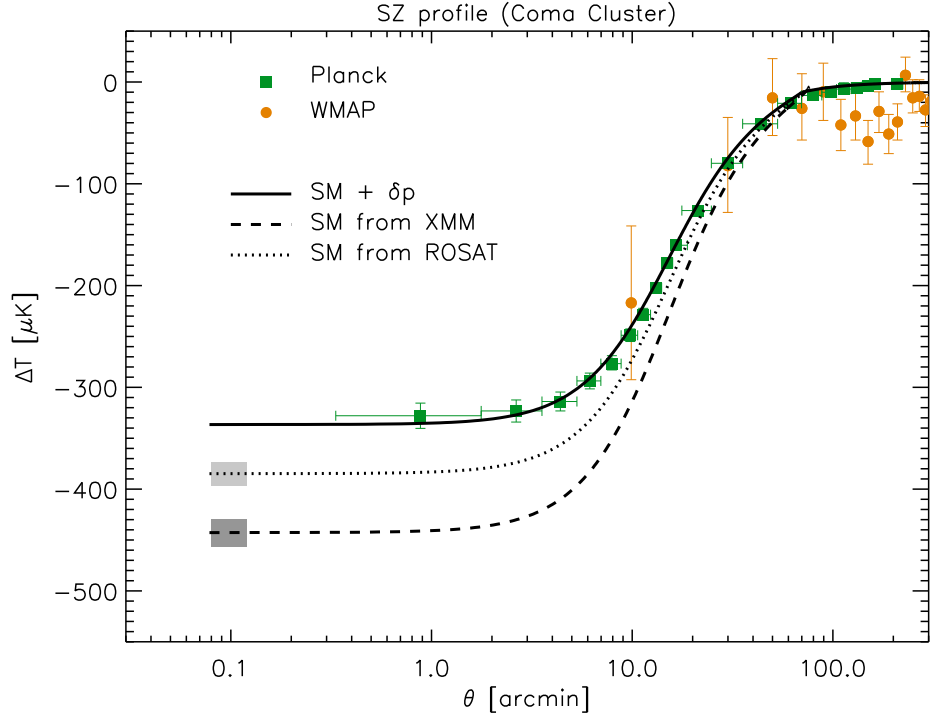}
\caption{Profiles of SZ effect toward the Coma Cluster. Green squares refer to
the \textsl{Planck} data by Ade et al. (2013), and orange circles to the
\textsl{WMAP} data by Komatsu et al. (2011). Dashed line illustrates the
Supermodel outcome (smoothed on the \textsl{Planck} resolution scale) based
on the fit to the X-ray data from \textsl{XMM-Newton}, with the heavy shaded
area representing the associated $2$-$\sigma$ uncertainty; dotted line and
light shaded area illustrate the same when basing on the fit to the X-ray
brightness from \textsl{ROSAT} data. The solid line is the outcome when a
non-thermal contribution $\delta\approx 20\%$ is included in the SM (see
Eq.~[12]), adding to the thermal X-ray pressure from \textsl{XMM-Newton} (or
$\delta\approx 15\%$ for \textsl{ROSAT}).}
\end{figure}

\clearpage
\begin{figure}
\center
\includegraphics[height=10cm]{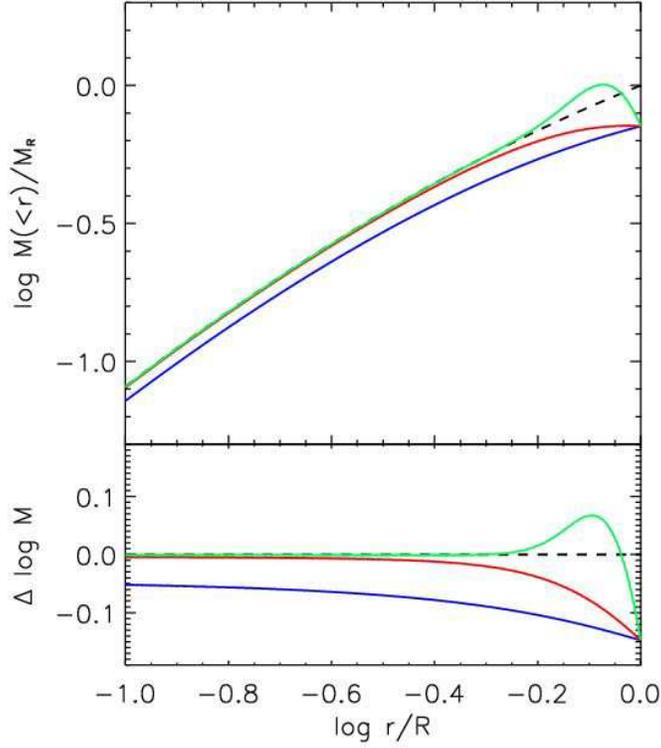}
\caption{Mock profiles of DM mass reconstructed from X rays. The dashed line
illustrates the input (true) mass, or equivalently that recovered on
including the non thermal component $\delta\equiv \delta p/p$; the solid
lines illustrate that reconstructed from X-ray observables on ignoring the
latter component (i.e., assuming strictly thermal hydrostatic equilibrium).
For $\delta$ we have adopted a turbulence profile
$\delta(r)=\delta_R\,\exp[-(R-r)^2/\tilde \ell^2]$ with $\delta_R=20\%$ at
the outer boundary, and different values of the decay scale $\tilde
\ell=0.9\, R$ (blue line), $\tilde 0.5\, R$ (red line) and $\tilde 0.2\, R$
(green line), see \S~5.1 for details.}
\end{figure}

\clearpage
\begin{table}[t]
\caption{A cluster library.} \center
\begin{tabular}{lccccc}
\hline\hline
Cluster & $z$ & Class & $k_c$ &  $a$ & $c$ \\
\hline
A1795   & 0.06 & LE & $15$ & $0$ & $8.5$\\
PKS0745 & 0.10 & LE & $15$ & $0$ & $7.6$ \\
A2204   & 0.15 & LE & $10$ & $0.16$ & $5.5$\\
A1413   & 0.14 & LE & $10$ & $0.36$ & $8.3$\\
A2597   & 0.08 & LE & $6$ & $0.71$ & $7.2$\\
A2199   & 0.03 & LE & $13$ & $0.95$ & $6.7$\\
A1689   & 0.18 & LE & $105$ & $0.80$ & $\sim 10$\\
\\
A2218   & 0.18 & HE & $350$ & $0.8$ & $5.1$\\
A399    & 0.07 & HE & $330$ & $1.0$ & $3.1$\\
A1656   & 0.02 & HE & $540$ & $1.3$ & $3.0$\\
A644    & 0.07 & HE & $124$ & $1.1$ & $3.9$\\
A2256   & 0.06 & HE & $248$ & $1.5$ & $2.7$\\
\hline
\end{tabular}
\end{table}
\noindent For further details on the fits, including uncertainties of the
parameter determinations and specific $\chi^2$ values (generally of order
$1$), see the extended Table 1 in Cavaliere et al. (2011a).

\end{document}